\documentclass[aps,pra,twocolumn,superscriptaddress,floatfix]{revtex4-2}

\usepackage{graphicx}   
\usepackage{amsmath}    
\usepackage{amssymb}    
\usepackage{bm}         
\usepackage{hyperref}   
\usepackage{physics}    
\usepackage{xcolor}     
\usepackage{float}
\begin{document}

\title{Approximate vortex lattices of atomic Fermi superfluid on a spherical surface}

\author{Keshab Sony}
\affiliation{Department of Physics, University of California, Merced, California 95343, USA}

\author{Yan He}
\affiliation{College of Physics, Sichuan University, Chengdu, Sichuan 610064, China}

\author{Chih-Chun Chien}
\email{cchien5@ucmerced.edu}
\affiliation{Department of Physics, University of California, Merced, California 95343, USA}

\date{\today}

\begin{abstract}
While planar Fermi superfluids form Abrikosov vortex lattices under magnetic or effective gauge fields, spherical geometry forbids perfect lattices above 20 vortices. We characterize approximate vortex lattices of atomic Fermi superfluids under an effective monopole field on a spherical surface as an analogue of the planar vortex-lattice problem by two constructions based on the Ginzburg-Landau theory. The first one is geometric and uses approximate lattices, including the random, geodesic-dome, and Fibonacci lattices, as scaffolds to construct the vortices from the degenerate monopole harmonics. The second one generalizes J. Phys. A 39, 2499 (1996) and minimizes the free energy by numerically adjusting the coefficients to find the solution with the minimal Abrikosov parameter. We have verified that the vortices from both constructions are zeros of the order parameter with circulating currents around the vortex cores and the Poincar\'e--Hopf
theorem is satisfied. As the number of vortices increases, the Abrikosov parameters of both the Fibonacci-lattice and minimization solutions extrapolate to the planar value. We briefly discuss implications for ultracold atoms in thin spherical-shell geometry.
\end{abstract}

\maketitle

\section{Introduction}
Ultracold atomic clouds in thin spherical shells have been realized via bubble traps in the International Space Station \cite{Carollo2022, Lundblad_2023} and multi-species phase-separation structures in spherical harmonic traps on earth \cite{PhysRevLett.97.060403, PhysRevLett.129.243402}. The spherical geometry has posed as a platform to test and characterize features of interacting quantum systems in curved space (see Refs.~\cite{articlenew,TononiReview23,TONONI20241} for a review). For example, quantum vortices as topological excitations may exhibit interesting behavior in curved space~\cite{RevModPhys.82.1301}. There have been theoretical works on vortices in Bose-Einstein condensates (BECs) on spherical surfaces \cite{PhysRevA.103.053306, PhysRevA.109.013301,PhysRevA.111.033322, PhysRevA.103.053306, PhysRevA.102.043305, PhysRevA.85.053631,10.1116/5.0211426,Arazo_2021}. Meanwhile, Ref. ~\cite{He22Sphere} derived the BCS theory in thin-shell geometry while Refs.~\cite{PhysRevB.77.024512,He23Vortex} have studied isolated vortices with higher angular momentum of Fermi superfluid on a spherical surface. In planar geometry, the Abrikosov vortex lattice~\cite{Abrikosov1957JETP,JonesMarch_TSSP2_Dover1985, PhysRev.133.A1226} emerges when a two-dimensional (2D) Fermi superfluid is subject to a strong magnetic or gauge field. It is an interesting and important questions on the vortex-lattice structure when Fermi superfluid is confined in a thin spherical shell. Previous studies of bosonic superfluids using the Gross-Pitaevskii equation~\cite{PhysRevA.102.043305, PhysRevA.109.013301} start to show multi-vortices configurations at high flux. Here we set out to identify and characterize spherical vortex lattices for atomic Fermi superfluids.

A fundamental mathematical theorem rules out any regular polyhedron with more than $20$ vertices~\cite{Cromwell1997}, thereby prohibiting the construction of perfect lattices with more than $20$ sites on a spherical surface having equal distance between nearest neighbors. 
Nevertheless, constructing an optimal lattice-like structure over a curved surface has gained significant attention in various areas of condensed matter physics. One example is the Thomson problem of finding the minimum energy configuration of classical charges on a sphere \cite{PhysRevLett.78.2681}. Studies of the generalized Thomson problem have shown that curvature induces topological defects and grain-boundary scars on a sphere~\cite{PhysRevLett.89.185502, article6}. Another example is the spherical formulation of the fractional quantum Hall effect (FQHE)~\cite{PhysRevLett.51.605}, leading to a hierarchy of states and topological quantities on a sphere \cite{PhysRevB.87.195131}. 
In a similar vein, here we will study possible vortex lattices of Fermi superfluids on a spherical surface and extend the well-established theory of the Abrikosov vortex lattices in planar geometry to investigate geometric influence to the vortex-lattice structures.

Since the Ginzburg-Landau (GL) theory (see Refs.~\cite{FetterWalecka2003,Mazenko2003FOD} for a review) of Fermi superfluid subject to a perpendicular magnetic field is behind the Arbikosov vortex lattice on a 2D plane, we will adopt a similar approach to investigate the vortex-lattice structures of Fermi superfluid confined to a spherical surface with an effective magnetic monopole at the center. The radial flux from the monopole plays the role of the perpendicular magnetic field in the planar case. The magnetic monopole proposed by Dirac \cite{Dirac1931} has inspired both theoretical proposals \cite{PhysRevLett.103.030401, PhysRevLett.120.130402} and experimental realizations \cite{Ray_2014, PhysRevX.7.021023} in cold-atom systems using synthetic gauge fields. The linearized GL equation gives the degenerate ground states for constructing the vortex-lattice solution. However, the spherical surface is not compatible with discrete translational symmetry, thereby prohibiting the original Abrikosov construction by tiling the solutions by discrete translation.
In contrast to the monopole field, we mention that for a unidirectional magnetic field, there is no degeneracy in the ground state of the linearized GL equation on a spherical surface~\cite{article3}, and the broken spherical symmetry under a unidirectional magnetic field leads Ref.~\cite{Du05} to develop a numerical construction of only a few vortices on a spherical surface while Ref.~\cite{Simonucci2015} shows arrays of vortices in three-dimensional cases.

The degenerate ground states of the linearized GL equation on a sphere are the monopole harmonics~\cite{Wu1976}, which will be used to construct possible vortex lattices on a spherical surface by using two different approaches. The first one is a geometric construction, which employs scaffolds from the random, geodesic-dome, and Fibonacci lattices to construct the corresponding vortex lattices on a sphere. This method places the zeros of the wavefunction at the scaffold. In the second approach, we employ numerical minimization to find the solution with the minimal Abrikosov parameter $\beta_A$, which reflects the non-uniformity of the solution. The minimization construction was first formulated in the seminal work~\cite{1996JPhA...29.2499D, PhysRevB.55.3816}, showing magic numbers of vortices corresponding to specific geometric-dome lattice structures.
We generalize the minimization construction to find approximate vortex lattices on a spherical sphere and compare the results with those from the geometric construction. Furthermore, we will check if the gauge-invariant current circulate around each vortex to confirm the topological property and show how the Poincar\'e--Hopf
theorem \cite{Hopf1927, MilnorTopology} is satisfied by both geometric and minimization constructions.

We also compare $\beta_A$ from different constructions. For smaller vortex numbers less than $20$, the geodesic-dome lattice yields the smallest $\beta_A$ since perfect lattices are still allowed. As the vortex count increases, the geometric construction using the Fibonacci lattice produces $\beta_A$ comparable to that of the minimization solution while both extrapolate to the same value $\beta_A\approx1.16$ given by the 2D planar triangular vortex lattice \cite{PhysRev.133.A1226, JonesMarch_TSSP2_Dover1985}. The comparison between the geometric and minimization constructions also indicates that the Fibonacci lattice provides a reasonable approximation of the spherical vortex lattice, which helps us to identify and characterize the structures in an analytic fashion. We caution that different from the Thomson problem and FQHE where minimizing the Coulomb interaction determines the structures is sought, the spherical vortex lattices studied here are influenced by the nonlinear interactions of the order parameter in the GL free energy.

The rest of the paper is structured as follow. Sec.~\ref{Sec:Theory} sets up the GL theory on a spherical surface and describes the geometric and minimization constructions of spherical vortex lattices by the degenerate ground states of the linearized GL equation. We also briefly review approximate lattice structures on a sphere. Sec.~\ref{Sec:Result} presents approximate vortex lattices on a sphere from the geometric constructions using the random, geodesic-dome, and Fibonacci lattices. The minimization solutions are shown to extrapolate to the same value as the Fibonacci lattice. Sec.~\ref{Sec:Implication} discusses possible implications of approximate spherical vortex lattices in ultracold atoms. Finally, Sec.~\ref{Sec:Conclusion} concludes our work.

\section{Theoretical Framework}\label{Sec:Theory}
\subsection{Ginzburg-Landau theory on a sphere}
We consider an equal-population Fermi superfluid confined to a spherical surface with an effective monopole of strength $g$ at the center of the sphere. 
The effective magnetic field $\mathbf{H(r)}=\frac{g}{r^2}\mathbf{\hat{r}}$ created by the monopole is perpendicular to the spherical surface at every point. Near the critical point when the effective magnetic field is strong and the order parameter is relatively small, the system is described by the Ginzburg-Landau (GL) theory~\cite{JonesMarch_TSSP2_Dover1985, Mazenko2003FOD} with the free energy functional given by 
\begin{equation}
 F_{\mathrm{2D}}[\psi] \;\simeq\;
     \int_{S^2} d^2r\;
    \left[
        \alpha\,|\psi|^2
        + \frac{\beta}{2}\,|\psi|^4
        + \frac{1}{2MR^2}\, {\psi^*} L^2 \psi
    \right]
\label{eq:GL_free_energy}
\end{equation} 
Here the complex Cooper-pair wavefunction is $\psi (\theta, \phi)=|\psi|e^{i\Phi}$, $M=2m_F$ is the Cooper-pair mass (with $m_F$ being the mass of a single fermion), and $L^2 = \mathbf{L\cdot L}$ with $\mathbf{L}=\mathbf{r} \cross (\mathbf{p}-(q/c)\mathbf{A})$ denoting the orbital angular momentum operator. We assume each Cooper pair carries an effective electric charge $q$ and will explain the vector potential $\mathbf{A}$ shortly. For charge-neutral Fermi superfluid, the effective charge represents the coupling strength to artificial gauge fields, such as laser-atom interactions \cite{Goldman2013LightinducedGF, RevModPhys.83.1523, articleultracold,PhysRevLett.103.030401}. The coefficients $\alpha$ and $\beta$ are functions of temperature and magnetic field, which can be determined from the microscopic theory~\cite{FetterWalecka2003}.

We rewrite the theory in dimensionless form by measuring the length in units of the shell radius \(R\), so that the area element becomes \(d^2r=R^2 d\Omega\), with \(d\Omega=\sin\theta\,d\theta\,d\phi\). In the following, energies are measured in units of \(E_0=\hbar^2/(2MR^2)\), the order parameter in units of the Ginzburg--Landau amplitude
\(\psi_0=\sqrt{|\alpha|/\beta}\) in the absence of the vector potential, the vector potential in units of
\(A_0=\hbar c/(qR)\), and the angular momentum in units of \(\hbar\).
We use the same symbols for the rescaled dimensionless quantities
hereafter. Near the upper critical field, the order parameter is small, \(|\psi|\ll 1\) (in units of $\psi_0$), and the quartic term in the Ginzburg-Landau free energy may be neglected.
The dimensionless linearized GL equation associated with the free energy functional can then be cast as an 
eigenvalue equation 
\begin{equation}
    H_{ang}\psi = E \psi,
    \label{eq:se}
\end{equation}
where $H_{ang}=L^2$ and $E=|\alpha|/E_0$. Here, $\alpha$ is related to the upper critical field $H_{c2}$ \cite{Tinkham1996, k2yd-vpbn}.

As pointed out in Ref.~\cite{Wu1976}, the modified angular momentum  
$\mathbf{J} = \mathbf{L} - \tilde{g}\mathbf{\hat{r}}$, where $\tilde{g} = qg/\hbar c$ is the dimensionless monopole magnetic charge,
satisfies the angular-momentum algebra 
$[J_\mu, J_\nu] = i  \epsilon_{\mu\nu\xi} J_\xi$, where $\mu,\nu,\xi=x,y,z$.
Moreover, $[J^2, J_z]=0$, allowing us to construct simultaneous eigenstates characterized by the quantum numbers $j$ and $m$. Explicitly, $J^2 |j, m\rangle = j(j+1) |j, m\rangle$ and $J_z |j, m\rangle = m |j, m\rangle$.
The quantum number $j=\tilde{g}, \tilde{g}+1, \cdots$ labels the Landau levels. For a given $j$, $m=-j, ....j$ represents the degenerate states.
Consequently, Eq.~\eqref{eq:se} has the form of the Schrödinger equation of an electron in a monopole field with $H_{ang}=J^2-\tilde{g}^2$. 
Ref.~\cite{Tamm1931} introduced the solution as generalized spherical harmonics, and Ref.~\cite{Wu1976} analyzed the problem via sections of the monopole bundle and defined the vector potentials in two regions to handle the singular behavior caused by the monopole. 
The simultaneous eigenstates are the monopole harmonics $Y_{\tilde{g}, j, m}(\theta, \phi)$~\cite{Wu1976}. 
Here we focus on the ground state at zero temperature, so we will take $j=\tilde{g}$ in the following.
For a monopole with effective magnetic charge $g$, the total flux through the surface is $4\pi g$. The flux has been shown to be quantized in units of the flux quantum~\cite{MQM}.
Assigning $q=2e$ to the effective electric charge carriers of the Cooper pairs, the Dirac quantization condition gives $\tilde{g}=2eg/\hbar c=N$ with integer values of $N$. Moreover, the effective flux of a vortex in the Fermi superfluid is $\phi_v=2\pi \hbar c/q$, so the total number of flux quanta on the sphere becomes
\begin{equation}
    N_v=\frac{4\pi g}{\phi_v}=2N.
\end{equation}

On a spherical surface, the vector potentials  from the monopole on $\mathcal R^{a,b}$ excluding the south or north pole are given by~\cite{Wu1976}
\begin{equation}\label{Eq:TwoA}
\begin{aligned}
    \textbf{A}^a &= \tilde{g} \frac{1 - \cos \theta}{ \sin \theta} \hat{\phi}, \quad \mathcal R^a: 0 \leq \theta < \pi, \\
    \textbf{A}^b &= -\tilde{g} \frac{1 + \cos \theta}{ \sin \theta} \hat{\phi}, \quad \mathcal R^b:  0 < \theta \leq \pi.
\end{aligned}
\end{equation} 
For the case with $\tilde{g} = j =N$ discussed here, the monopole harmonics on $\mathcal R^a$ takes the form
\begin{equation}
    Y_{N,N,m}(\theta, \phi) = K_{N,m} \left( \sin\frac{\theta}{2} \right)^{N+m} \left( \cos\frac{\theta}{2} \right)^{N-m} e^{i(N+m)\phi},
    \label{eq:mh}
\end{equation}
where $K_{N,m} = (-1)^{N+m} \left[ \frac{2N+1}{4\pi} \frac{(2N)!}{(N+m)!(N-m)!} \right]^{1/2}$.
On $\mathcal R^b$, one may choose
$Y_{N,N,m}^b= Y_{N,N,m} e^{-2iN\phi}$~\cite{Wu1976}. While the monopole harmonics are defined in specific regions (excluding either the north or south pole), the order-parameter density $|\psi|^2$ is gauge-invariant and remains unaffected by the choice of regions. It suffices to use Eq.~\eqref{eq:mh} to construct the solution on
$\mathcal R^a$ since only one point (the south pole) is excluded. Nevertheless, a careful examination of the corresponding quantities on $\mathcal R^b$ shows that gauge-invariant physical quantities,
such as $|\psi|^2$ and the current, are the same in the overlapped region while patching the missing value at the south pole. 
To create a vortex-lattice solution, we superpose the monopole harmonics states with different $m$ according to 
\begin{equation}\label{Eq:Wavefunction}
    \psi(\theta,\phi)=\sum_{m=-N}^N C_m Y_{N,N,m}(\theta,\phi),
\end{equation}
where the $2N+1$ coefficients $C_m$ are undetermined. Each vortex corresponds to a zero of $\psi(\theta, \phi)$. 

A quantum vortex is associated with a circulating supercurrent around the vortex core. In Ref.~\cite{1996JPhA...29.2499D}, the vortex character of the zeros was identified through the $2\pi$ phase winding of the order parameter around them. In the presence of the monopole field, we complement the criterion by explicitly calculating the gauge-invariant current density~\cite{MQM,FetterWalecka2003}. Explicitly,
\begin{equation}\label{eq:current}
\mathbf{j}_s  
= \frac{1}{2i}(\psi^* \nabla \psi - \psi \nabla \psi^*)
- |\psi|^2 \mathbf{A}.
\end{equation}
It can be verified that by using the matching set of $\mathbf{A}^{a,b}$ and monopole harmonics on $\mathcal R^{a,b}$, the same current density is obtained consistently. 
We will show that the current circulates around the vortex core and provides a measurable quantity for detecting the vortex.

To determine the stability of a vortex-lattice solution \(\psi(\mathbf{r})\), it is crucial to check the Abrikosov parameter \cite{Abrikosov1957JETP,JonesMarch_TSSP2_Dover1985, PhysRev.133.A1226} 
\begin{equation}\label{Eq:beta}
\beta_A = \frac{\langle |\psi|^4 \rangle}{\langle |\psi|^2 \rangle^2}.
\end{equation}
Here 
\(\langle f\rangle=(4\pi)^{-1}\int_{S^2}f(\theta,\phi)d\Omega\) denotes the average over the spherical surface.
Since the nonlinear term increases the GL free energy, \(\beta_A\) reflects how evenly the vortices are distributed to minimize the overall energy. In Abrikosov's original work \cite{Abrikosov1957JETP}, the square lattice of vortices gives $\beta_A\simeq 1.18$. Later numerical evaluations on a 2D plane with periodic boundary condition showed
the triangular lattice has $\beta_A \simeq 1.16$ \cite{PhysRev.133.A1226}. Therefore, the triangular vortex lattice is energetically more favorable than the square vortex lattice, despite their extremely close values of $\beta_A$.

\subsection{Geometric construction of spherical vortex lattices}
Although a perfect lattice with more than $20$ vortices with equal distance between nearest neighbors cannot form on a spherical surface due to the lack of regular polyhedra above the limit, there have been approximate spherical lattices in mathematics. For example, the random lattices with their vortices randomly placed on the sphere offer a class of possible structures, despite their lack of local uniformity in each realization. We also explore two deterministic approximations, the Fibonacci and geodesic-dome lattices. In the following, we will present a geometric construction for approximate spherical lattices, exemplified by the random, Fibonacci, and geodesic-dome lattices.

The random lattice draws random points from a uniform distribution over the surface of a sphere in one realization. Two independent random variables, \(u_1, u_2 \in [0,1]\), are generated, and the corresponding polar coordinates are computed as follows.
\begin{equation}
\phi = 2\pi u_1, \quad \theta = \arccos(2u_2 - 1),
\end{equation}
where \( \phi \in [0, 2\pi) \) is the azimuthal angle and \( \theta \in [0, \pi] \) is the polar angle \cite{tashiro1977methods}. This method ensures a uniform distribution of points in \( \cos \theta \) and \( \phi \) since the differential solid angle has the form \( d\Omega = \sin \theta \, d\theta \, d\phi = -d(\cos \theta) \, d\phi \). The resulting points are randomly distributed and uncorrelated, providing a reference for comparison with deterministic lattice structures.

The counterpart of the triangular lattice on a sphere may be constructed by the icosahedral geodesic-dome lattice derived by projecting a specific class of polyhedra, known as icosadeltahedra characterized by their equilateral triangular faces and icosahedral symmetry, onto the surface of a sphere \cite{WenningerSphericalModelsDover2012}. One begins with a regular icosahedron, where each of its 20 triangular faces is subdivided, resulting in a grid that maintains the symmetry of the icosahedron while covering the spherical surface. The subdivision is parameterized by two non-negative integers $(h,k)$ that count the steps along the triangular lattice vectors on each icosahedral face. The triangulation number
$T = h^2 + hk + k^2 \quad (T = 1, 3, 4, 7, 9, 12, 13, 16, 19, \dots)
$
gives the number of smaller equilateral triangles per icosahedral face. The total number of faces is $20T$, leading to
\begin{equation}
N_v = 10T + 2
\end{equation}
vertices on the sphere.
Three distinct geodesic-dome classes arise from specific choices of $(h,k)$ \cite{WenningerSphericalModelsDover2012}: 
\begin{itemize}
    \item \textbf{Class I} $(h,0)$: edge-aligned subdivision ($T = h^2$).
    \item \textbf{Class II} $(h,h)$: triacon subdivision with $60^\circ$ rotation ($T = 3h^2$).
    \item \textbf{Class III} $(h,k)$ with $h>k>0$: general asymmetric subdivision.
\end{itemize}
We caution that the geodesic-dome lattices contain disclinations, i.e., vertices with only five nearest neighbors instead of six for the regular triangular lattice. The existence of the disclinations are necessary to make the geodesic-dome lattices fit on the spherical surface.

The Fibonacci lattice \cite{Santos2013} provides an analytical construction of $N_v$ approximately uniformly spaced points on a sphere. For $n = 1,\dots,N_v$, the spherical coordinates of the Fibonacci lattice are 
\begin{equation}
\theta_n = \arccos\!\left(\frac{2n-1}{N_v}-1\right),~
\phi_n = 2\pi n \varphi^{-1} \pmod{2\pi},
\end{equation}
where $\varphi = (1+\sqrt{5})/2$ is the golden ratio. 
Different from the geodesic-dome lattices which can only take specific values of the total vertex number $N_v$, the Fibonacci lattices can be constructed for arbitrary positive integer $N_v$. This property of the Fibonacci lattice allows for a detailed analysis of its behavior as $N_v$ increases. We mention that vortices of the classical XY model has been studied in the Fibonacci lattice~\cite{PhysRevResearch.4.023005}.

To construct approximate vortex lattices using the random, geodesic-dome, and Fibonacci lattices as scaffolds on a spherical surface, we will build
a wave-function from the monopole harmonics with the prescribed zeros located at the scaffold. Explicitly, we solve the following equations with $(\theta_n,\phi_n)$ determined by the random, geodesic-dome, and Fibonacci lattices.
\begin{equation}
    \sum_{m=-N}^N C_m Y_{N,N,m}(\theta_n,\phi_n)=0,\qquad
n=1,\cdots,N_v
\label{eq:solu}
\end{equation}
Since $N_v=2N$, we have $2N$ equations with $2N+1$ unknowns. Therefore, a set of non-zero $C_m$ can be found.
Once the coefficients are determined, we get a wave-function $\psi(\theta,\phi)$ which have $2N$ zeros located at the prescribed lattice sites.

We caution that while the construction pinpoints the $N_v$ zeros of the wave-function, it does not establish whether the zeros are vortices or not. We will check the circulation of gauge-invariant currents around the zeros to confirm the solution from the geometric construction indeed has $N_v$ vortices.

\subsection{Minimization construction of spherical vortex lattices}
For a comparison, we also used a numerical optimization to find the coefficients $\{ C_m\}$ in Eq.~\eqref{eq:solu} that minimize the Abrikosov parameter $\beta_A$. This idea has been implemented in Ref.~\cite{1996JPhA...29.2499D}, which found a series of minimal-$\beta_A$ solutions for vortices on a spherical surface. Interestingly, the "magic numbers" of vortices with lower energies are from structures resembling the geometric-dome lattices. However, only partial series of Class I, II and III of the geodesic-dome lattices were identified by Ref.~\cite{1996JPhA...29.2499D}. Moreover, the minimization only produces the order-parameter profile without explicitly showing the gauge-invariant current around the vortices.

Here we revisit the minimization method and compare the results with the geometric constructions. We perform the minimization by varying the real and imaginary parts of the complex coefficients in Eq.~\eqref{Eq:Wavefunction}, which constitutes a parameter space of \(2(2N+1)\) real dimensions. To this end, we employ two gradient-based optimization algorithms from the SciPy library: The L-BFGS-B method and a trust-region constrained approach. The L-BFGS-B is a quasi-Newton algorithm that approximates the Hessian matrix, enabling efficient handling of high-dimensional problems with bound constraints \cite{nocedal2006numerical}. The trust-region method constructs a local quadratic model of the objective function and adaptively adjusts the step size to ensure robust convergence in nonlinear settings \cite{yuan2000review}.

Convergence for both methods is determined by checking the relative change in the function value 
and the norm of the gradient.
To mitigate the risk of converging to local minima, we perform multiple optimization runs starting from random initial guesses, with initial amplitudes biased toward smaller \(|m|\) to enhance numerical conditioning. The solution yielding the lowest value of \(\beta_A\) is retained. Using the optimal coefficients \(C_m^\star\), we compute the corresponding density profile \(|\psi(\theta,\phi)|^2\) on a discretized \((\theta,\phi)\) grid and store it for subsequent analysis.

\section{Results}\label{Sec:Result}
\subsection{Geometric constructions}
\begin{figure}[t]
    \centering
\includegraphics[width=0.48\textwidth]{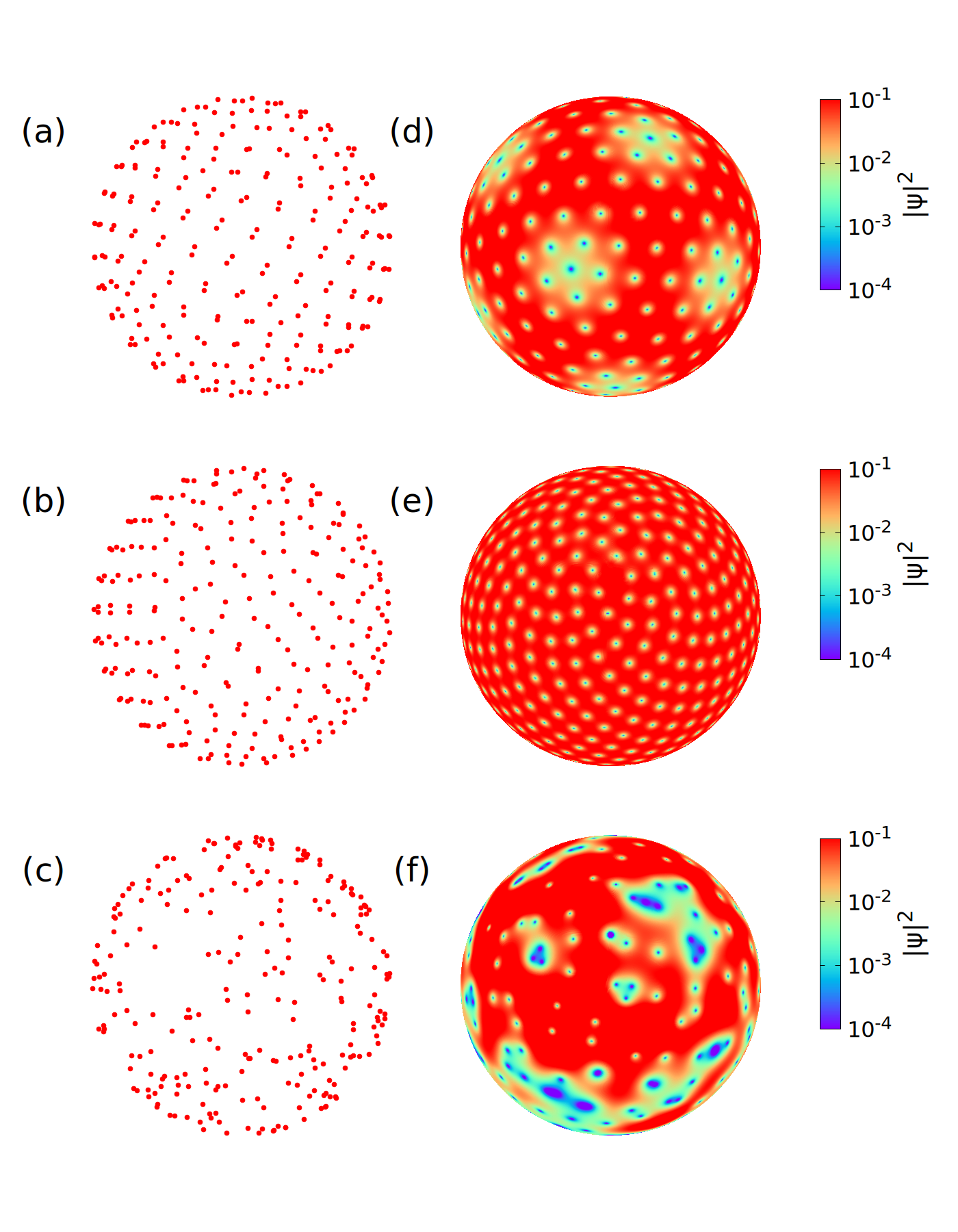}
    \caption{
    Left column: Illustrations of the (a) geodesic-dome, (b) Fibonacci, and (c) random lattices on the sphere 
    used as scaffolds for constructing spherical vortex lattices. 
    Right column: The probability distributions $|\psi|^2$ from the geometric constructions of the corresponding vortex lattices (d) - (f). The vortex cores are zeros of the wave function. Here $N_v=252$ for all panels. For the random lattices of (c) and (f), only one realization is presented.
    }
    \label{fig:lattices}
\end{figure}

We numerically construct the wavefunctions according to Eq.~\eqref{Eq:Wavefunction} with the targeted vortices satisfying Eq.~\eqref{eq:solu} with the scaffolds from the random, Fibonacci, and geodesic-dome lattices on a spherical surface. For each lattice type, the wavefunction from the geometric construction has exactly $N_v=2N$ zeros at the specified locations.
The left column of Figure~\ref{fig:lattices} illustrates the three scaffolds for the geometric construction.

The corresponding probability distributions $|\psi(\theta,\phi)|^2$ constructed by using the Fibonacci, geodesic-dome, and random lattices as scaffolds
are shown in the right column of Fig.~\ref{fig:lattices}. Each dark spot corresponds to a vortex core, where the order parameter vanishes. For the random-lattice result, we show one realization where clusters of nearby vortices may not be resolved at the available resolution. The vortex locations and patterns vary in different realizations. If averaged over many realizations, the probability returns to the uniform distribution, but this is not how we want to view the vortex structures.
For the geodesic-dome vortex lattice, one can see brighter spots around the sites with only five neighbors, indicating non-uniformity around the defects that makes an approximate lattice on a sphere possible. In contrast, the Fibonacci lattice tends to be more uniform, despite the existence of similar defects with less (or more) neighboring points.

\begin{figure}[t]
    \centering
    \includegraphics[width=0.8\linewidth]{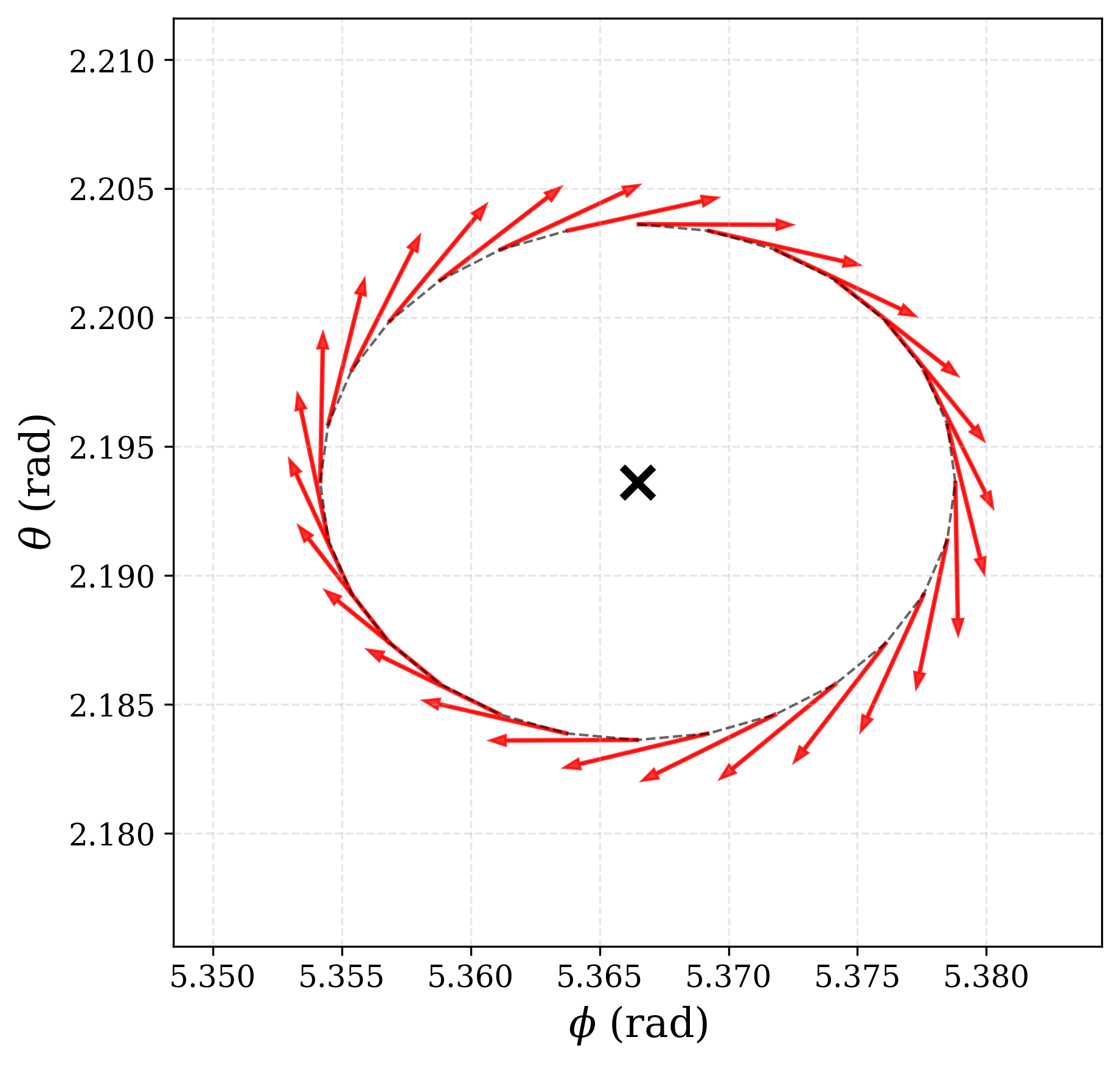}
    \caption{Current circulation (indicated by the arrows) around one of the vortices. The vortex core is labeled by the cross. The length of the arrow is proportional to the magnitude of the current.
    }
    \label{fig:current}
\end{figure}

The geometric construction only locates the zeros of the wave function. It is important to verify the topological nature of the vortex by checking the gauge-invariant current around each of the vortex. Fig.~\ref{fig:current} shows the current according to Eq.~\eqref{eq:current} around a closed loop enclosing a chosen vortex. The current indeed circulates around the vortex core and confirms the topological nature of the vortex.
Moreover, we found that all the vortices have the same circulation orientation in our solutions. Each vortex carries a $2\pi$ rotation of the gauge-invariant current. The sum of the magnetic flux from the vortices thus matches the total monopole flux due to its magnetic charge $N$. The spherical surface as a closed compact manifold thus provides an elegant example where the flux from the monopole escapes through the formation of vortices.
Similar current circulation can be observed around any vortex shown in the right column of Fig.~\ref{fig:lattices}, as well as vortices constructed by the minimization method presented later. We caution that for a loop close to a selected vortex core, the magnitude of the current is roughly constant while its orientation rotates around the vortex core by $2\pi$, as shown in Fig.~\ref{fig:current}. Away from the vortex core, nearby vortices affect the magnitude of the current, so both the magnitude and orientation can change along a larger loop around a vortex.

\begin{figure}[t]
    \centering
    \includegraphics[width=0.46\textwidth]{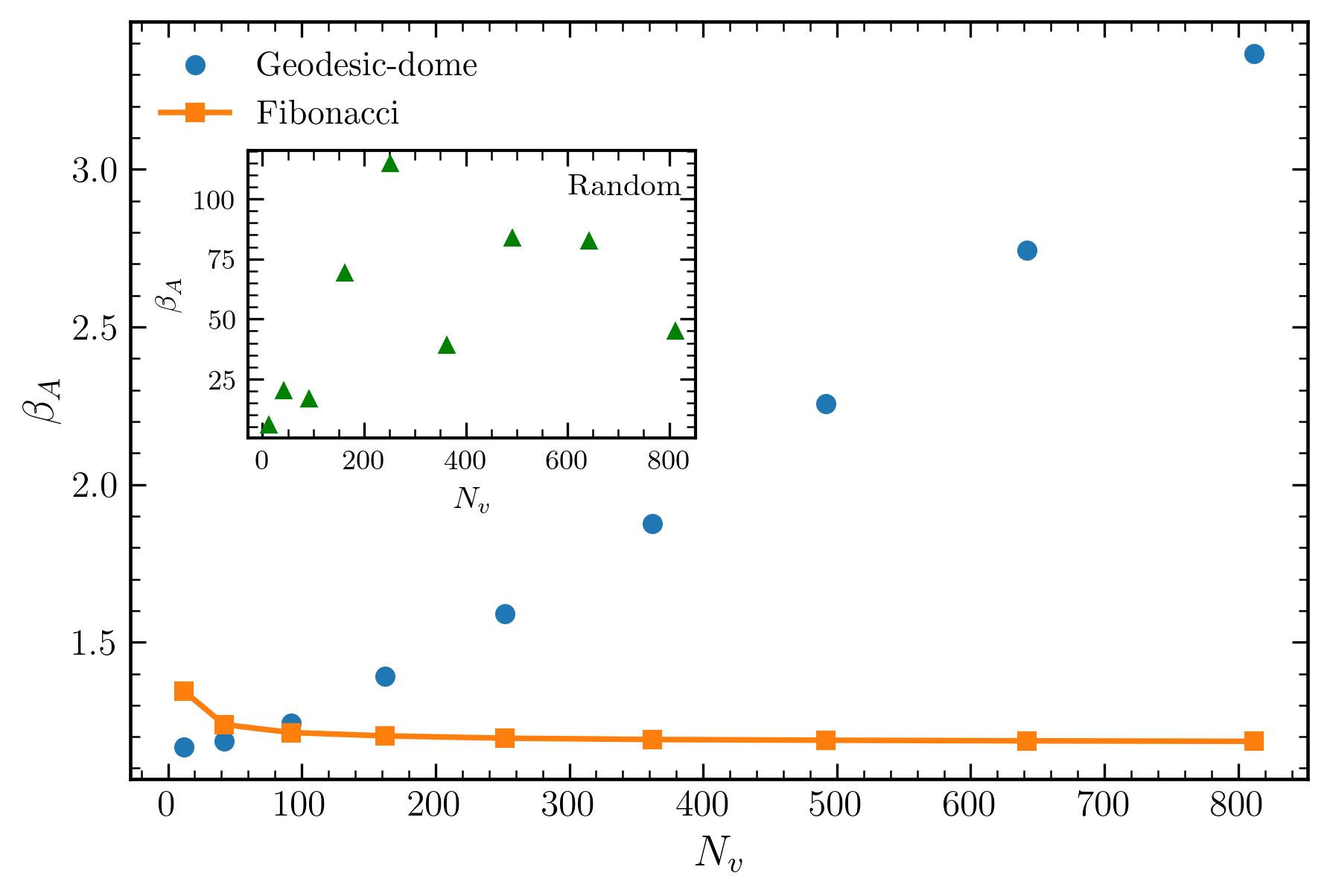}
    \caption{Abrikosov parameter $\beta_A$ as a function of the number of vortices $N_v$ from the geometric constructions using the Fibonacci (squares), geodesic-dome (circles), and random (triangles) lattices as scaffolds. While the construction works for the random and Fibonacci lattices with arbitrary positive $N_v$, only specific values are allowed for the geodesic-dome lattice.}
    \label{fig:energy}
\end{figure}

Figure~\ref{fig:energy} shows the values of the Abrikosov parameter $\beta_A$ for the three types of spherical vortices constructed from geometric scaffolds.
For small vortex numbers, the geodesic-dome prescription yields the smallest $\beta_A$. For example, $N_v=12$ corresponds to the icosahedron, which is a regular lattice. 
With increasing $N_v$, however, the Fibonacci lattice produces progressively smaller $\beta_A$ while the values of $\beta_A$ of the geodesic-dome lattices with specifically allowed vertex numbers increase rapidly. This is consistent with the illustration of Fig.~\ref{fig:lattices}, where the defects of the geodesic-dome lattice tend to have non-uniform probability distributions while the Fibonacci lattice is more uniform. Meanwhile, the random lattice from one realization results in large and highly fluctuating $\beta_A$, reflecting the strong inhomogeneity.


\subsection{Numerical minimization}
For a given $N_v$, a vortex-lattice structure can be obtained by minimizing $\beta_A$ with respect to the coefficients. As pointed out in Ref.~\cite{1996JPhA...29.2499D}, this method produced low-$\beta_A$ solutions at some of the geometric-dome lattices. It is unclear, however, what is the vortex structure in the large-$N_v$ limit. Our minimization constructions result in comparable low $\beta_A$ solutions but allow us to explore the large-$N_v$ limit. Moreover, we have checked each vortex indeed has current circulation similar to that shown in Fig.~\ref{fig:current}, which has not been clearly verified in the previous work.
We again emphasize that the gauge transformation of Eq.~\eqref{Eq:TwoA} with the corresponding monopole harmonics on a spherical surface lead to the same gauge-invariant quantities, such as the order-parameter density and current, in the overlapped region. This applies to both geometric and minimization constructions, and we have verified that using the set appropriate at the south pole does not produce any additional singular behavior.

\begin{figure}[t]
    \centering
    \includegraphics[width=1.0\linewidth]{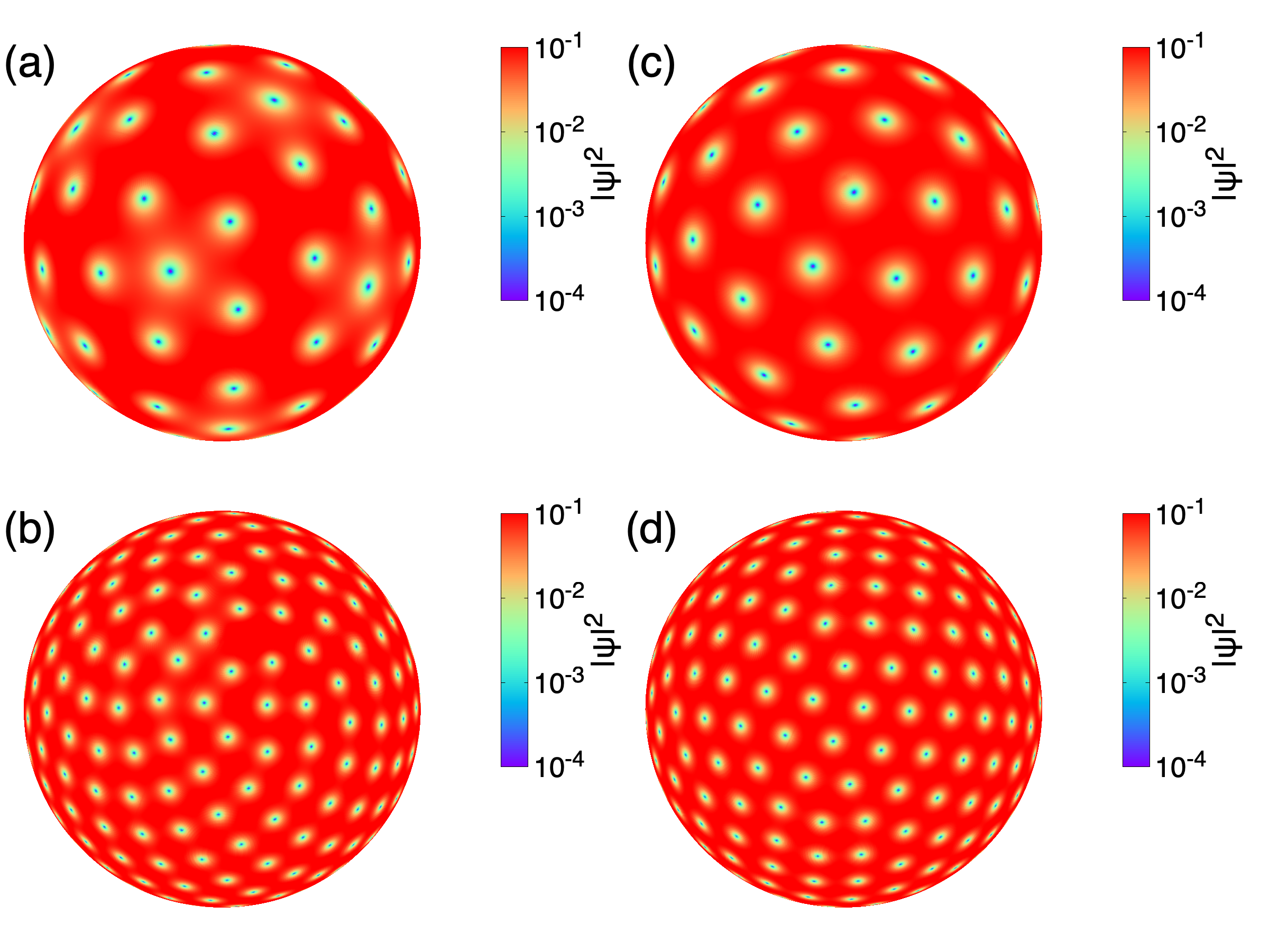}
    \caption{Comparison of geometrically constructions [(a) and (b)] and numerical minimization methods [(c) and (d)] of spherical vortex lattices.
        (a) Geodesic-dome lattice with \(N_{\text{v}} = 72\) and (b) Fibonacci lattice with \(N_{\text{v}} = 252\).
        (c) and (d) show the corresponding numbers of vortices obtained by minimizing \(\beta_{\text{A}}\) with respect to the coefficients.
        }
    \label{fig:compare}
\end{figure}

In Fig. \ref{fig:compare}, we compared the vortex lattices obtained from the geometrical construction with those from numerical minimization. For \( N_v = 72 \), the minimization leads to a lattice structure resembling the geodesic-dome lattice. A closer examination shows that the non-uniform probability distributions around the five-neighbor defects in the geometrical construction get smoothed out in the solution from minimization. As a consequence, the value of $\beta_A$ from minimization is typically lower than that from the geometric construction using the geodesic-dome lattice as a scaffold. For the $N_v=252$ cases shown in Fig. \ref{fig:compare}, $\beta_A\simeq 1.27$ for the geometric construction while $\beta_A\simeq 1.17$ for the minimization. This is also consistent with the finding of Ref.~\cite{1996JPhA...29.2499D} that the particular values of $N_v$ corresponding to the geometric-dome lattices may exhibit dips in the plot of $\beta_A$.

\begin{figure}[t]
    \centering
    \includegraphics[width=\linewidth]{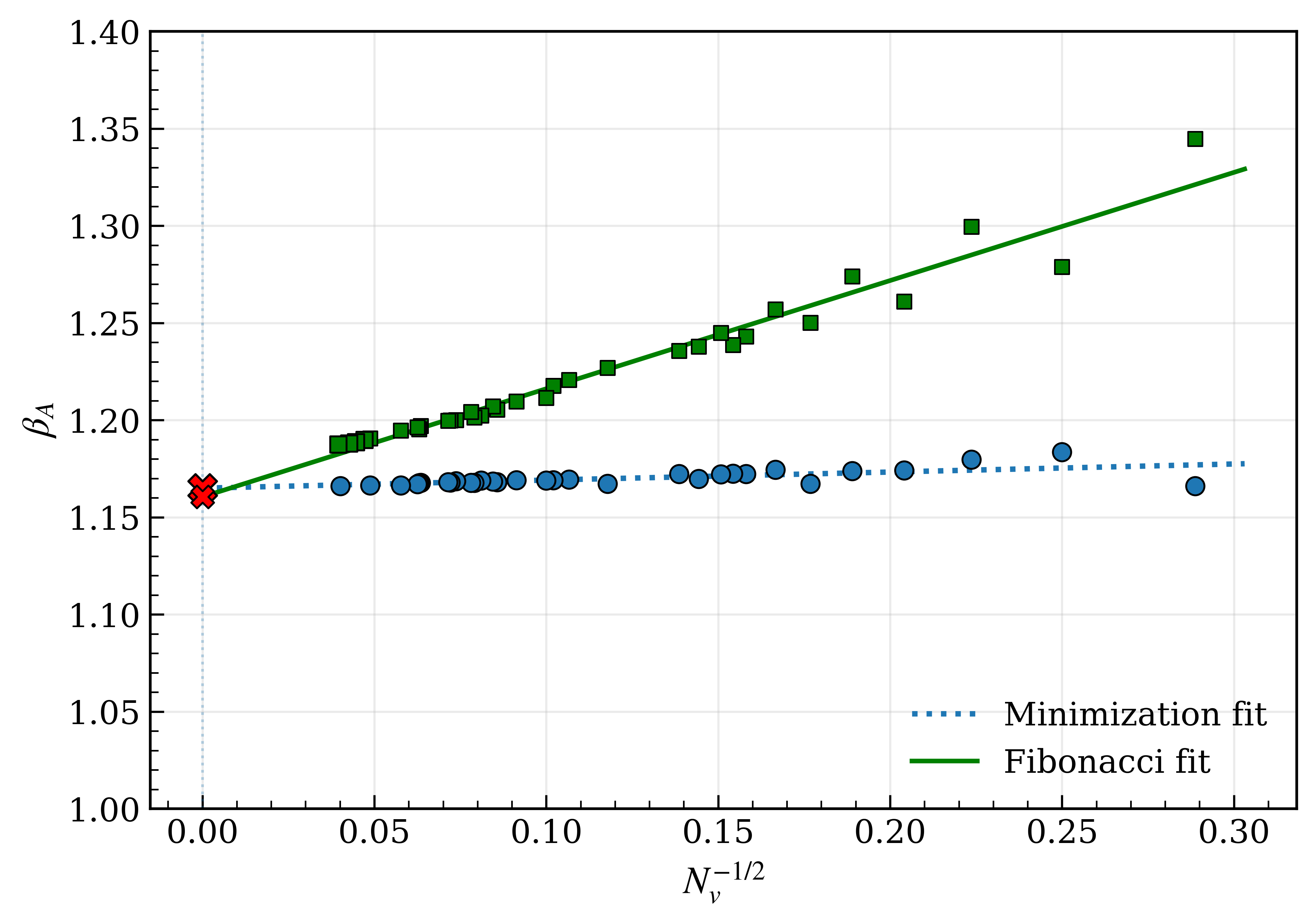}
    \caption{$\beta_A$ of the Fibonacci lattice (squares) and minimization solution (circles) as a function of ${N_v}^{-1/2}$. The extrapolations of both structures in the limit $N_v\rightarrow \infty$ converge to $1.16$ (the cross symbol).
    }
    \label{fig:extra}
\end{figure}

As $N_v$ increases, the solution from minimization starts to resemble that of the Fibonacci lattice. The bottom row of Fig. \ref{fig:compare} shows visually similar results for \( N_v = 252 \) from the geometric and minimization constructions. Fig.~\ref{fig:extra} shows the scaling behavior of $\beta_A$ vs $N_v^{-1/2}$, a scaling proposed in Ref.~\cite{1996JPhA...29.2499D}, for the geometric construction with the Fibonacci lattice and the minimization construction. The Fibonacci lattice has slightly higher $\beta_A$ than the minimization solution but approaches it as $N_v$ increases.
An extrapolation to $N_v \rightarrow \infty$ shows that both the Fibonacci lattice and minimization solution converge to the same value $\beta_A \approx 1.16$, which happens to be that for the triangular vortex lattices on a 2D plane \cite{Abrikosov1957JETP, JonesMarch_TSSP2_Dover1985}. This is understandable since in the infinite $N_v$ limit, the vortices are so dense that we may locally treat each region as a flat surface. The extrapolation also confirms the numerical accuracy of our geometric and minimization constructions. Furthermore, the Fibonacci lattice provides a manageable approximation to the fully numerical minimization on the spherical surface. The convergence of the extrapolations also suggests that, despite the lack of a genuine regular spherical lattice, approximate lattices manage to find an agreeable pattern in the large $N_v$ limit.



We remark that there are slight shifts in the vortex positions and spreading of the probability distribution from the numerical minimization solution relative to the corresponding geodesic-dome or Fibonacci lattices from the geometric construction. Therefore, the geometric construction provides computationally manageable and conceptually understandable approximations of spherical vortex lattices while the numerical minimization fine tunes the locations and weights of the vortices to lower $\beta_A$ further. When $N_v$ meets the specific numbers of the geodesic-dome structures, one expects the minimal-$\beta_A$ solution to resemble the geodesic-dome lattice. As $N_v$ increases, the Fibonacci lattice serves as a close approximation to the minimal-$\beta_A$ solution.

We comment on the value of the geometric construction despite the three examples (random, Fibonacci, and geodesic-dome) have higher $\beta_A$ than that of the minimization construction. First, the geometric construction only requires the solution to a set of linear equations, in contrast to the iterative search for the minimal solution involving nonlinearity. When the number of vortices increases, the approximate solutions from the geometric construction may serve as quick and efficient estimations. Second, mathematical developments of approximate lattices on a spherical surface are still ongoing. When new spherical lattice structures are revealed through mathematical discoveries, the geometric construction offers an immediate check of whether the minimization solution can be better approximated. Last but not least, the geometric construction provides some physical intuitions. For example, the magnetic charge of the monopole gives rise to the total magnetic flux, which in term determines the total number of vortices. While the geometric construction explicitly locates the vortices, it is less obvious how the minimization distributes the flux through the iterations.

\begin{figure}[t]
    \centering
    \includegraphics[width=1.0\linewidth]{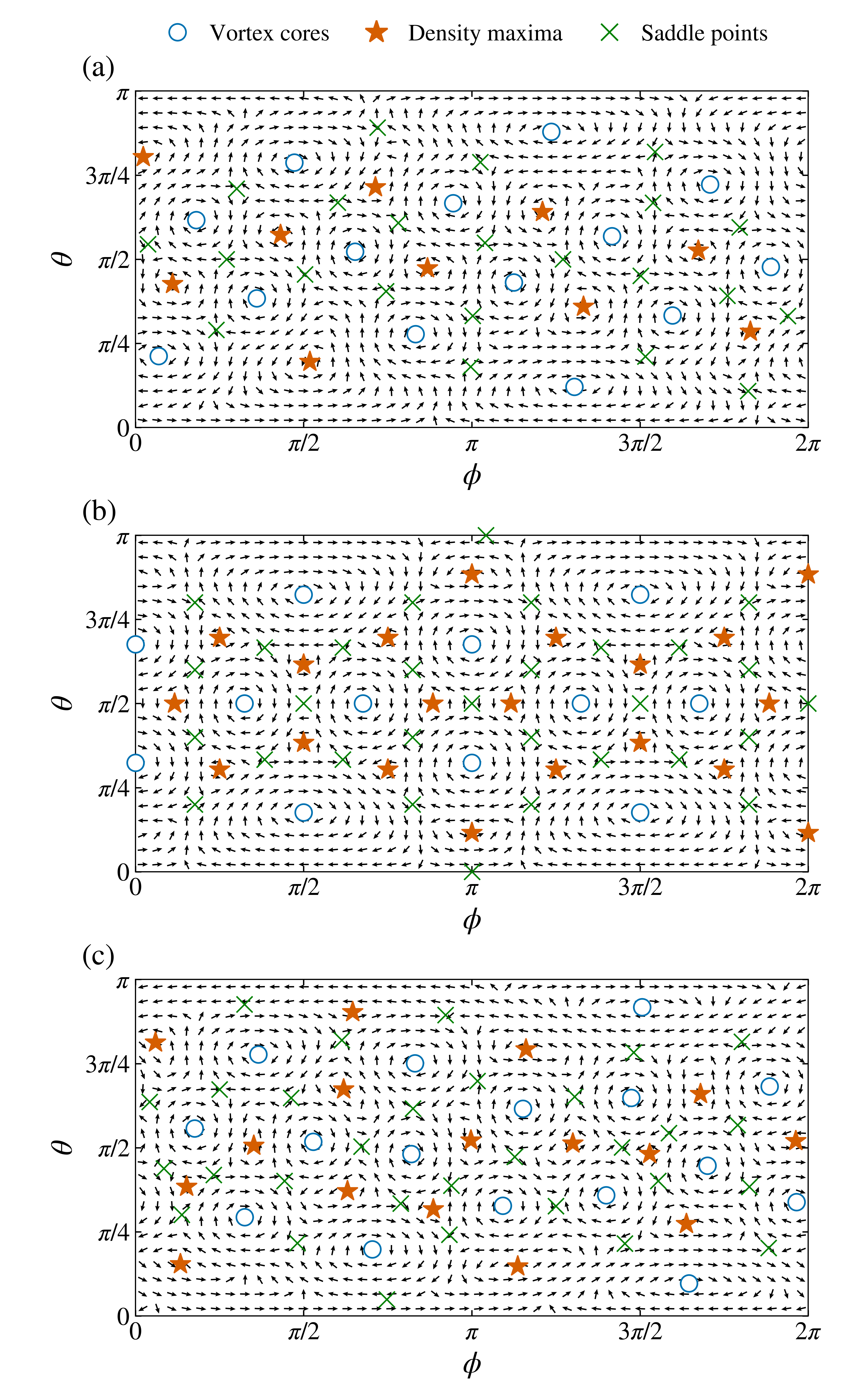}
    \caption{
Vector-field plots of the gauge-invariant current as a function of $(\phi,\theta)$ for (a) the Fibonacci lattice with $N_v=14$,
(b) the geodesic-dome lattice with $N_v=12$, and (c) the minimization solution with $N_v=16$. The circles, stars and crosses indicate the
vortex cores, maxima of the order-parameter density, and saddle-points. Here the vectors are normalized to the same length to emphasize their orientations.
}
    \label{fig:pht}
\end{figure}

We further examine the gauge-invariant current to verify the
Poincar\'e--Hopf theorem. Fig.~\ref{fig:pht} shows the vector-field plots of the current from the geometric and minimization constructions of approximate vortex lattices on a spherical surface. For a tangent vector
field $\mathbf{X}$ on a two-dimensional compact manifold \(M\) with isolated stagnation points, the Poincar\'e--Hopf
theorem \cite{Hopf1927, MilnorTopology} states that
\begin{equation}\label{Eq:PHT}
    \sum_p I_p (\mathbf{X})=\chi(M),
\end{equation}
where \(I_p\) is the
poincar\'e index of $\mathbf{X}$ at each stagnation point and \(\chi(M)\) is the Euler characteristic of the manifold. For the spherical surface considered here, \(M=S^2\), and hence $\chi(S^2)=2$. Here the relevant tangent vector field is the gauge-invariant current $\mathbf j_s$ shown in Eq~\eqref{eq:current}, and the stagnation points entering the
Poincar\'e--Hopf theorem are indicated by
\(\mathbf j_s=0\). To compute the Poincar\'e index at an isolated stagnation point with the coordinate 
\(p=(\theta_0,\phi_0)\), one may introduce the local tangent-plane coordinates
$x=(\sin\theta_0)(\phi-\phi_0), y=\theta-\theta_0$. 
In this local chart, the current of Eq.~\eqref{eq:current} has components \((j_x,j_y)\), and the
Poincar\'e index is given by~\cite{ArnoldODE, Gnther2021IntroductionTV}
\begin{equation}
    I_p=\frac{1}{2\pi}\oint_{C} d[\arg(j_x+i j_y)],
\end{equation}
where \(C\) is a small loop enclosing only the stagnation point at \(p\). From this definition, the vortex cores and maxima of the order-parameter density are center-type stagnation points with \(I_p=+1\) because the current-vector direction completes one counterclockwise rotation as a small loop around the point is traversed once in the counterclockwise direction, irrespective of the local circulation orientation. The additional stagnation points are the saddle-type points characterized by a hyperbolic current pattern. Around these saddle points, the current-vector direction completes one clockwise rotation along a counterclockwise loop, thereby giving rise to \(I_p=-1\) \cite{universe8040223}.

To better visualize the vector-field of the current, we normalize the arrows in Fig.~\ref{fig:pht} to the same length and focus on their orientations. For the Fibonacci lattice in Fig.~\ref{fig:pht}(a) with $N_v=14$, we identify 14
vortex-core stagnation points, 10  stagnation points associated with the maxima of the order-parameter density, and 22 saddle-type stagnation points. The corresponding index count is
$
14_{\rm vortex}+10_{\rm max}-22_{\rm saddle}=2
$.
For the geodesic-dome lattice in Fig.~\ref{fig:pht}(b) with $N_v=12$, the corresponding
counting leads to
$
12_{\rm vortex}+20_{\rm max}-30_{\rm saddle}=2$.
Finally, for the minimization solution with $N_v=16$ shown in Fig.~\ref{fig:pht}(c), 
$16_{\rm{vortex}}+16_{\rm{max}}-30_{\rm{saddle}}=2$.
Thus, after all stagnation points of the tangent current field are included correctly,
the Poincar\'e--Hopf theorem of Eq.~\eqref{Eq:PHT} is always satisfied.

\section{Implications}\label{Sec:Implication}
The magnetic monopole~\cite{Dirac1931}, despite a lack of evidence of its fundamental existence~\cite{PhysRevLett.124.031802}, has found a tangible analog in a synthetic setting using spinor BECs~\cite{Ray_2014,PhysRevX.7.021023}. 
As the spinor BEC adiabatically follows a sweep of local magnetic fields, it acquires a spatially varying texture that generates a superfluid velocity field of the monopole form. Since synthetic or artificial gauge fields can also be applied to atomic Fermi gases~\cite{RevModPhys.83.1523}, similar monopole fields may be generated for studying the spherical vortex lattices discussed here. However, we caution that spherical shells of atomic Fermi superfluid and high-field artificial monopoles still await experimental realizations, and their integration will be a challenging but rewarding task in the future.

Previous ultracold-atom experiments indicate that vortex cores are typically much smaller than the size of the atomic cloud \cite{doi:10.1126/science.1060182,Zwierlein2005VorticesAS}. The vortex-core size is set by the healing length \(\xi\)~\cite{FetterWalecka2003}. For atomic BEC, experimentally imaged core diameters are of the order of \(0.4\)–\(0.5~\mu\mathrm{m}\) \cite{doi:10.1126/science.1060182}. For fermionic superfluids, 
estimations of the vortex-core size in the BCS regime are around \(0.1~\mu\mathrm{m}\) \cite{Zwierlein2005VorticesAS}. Those vortex-core sizes can be much smaller than the characteristic sizes of shell-shaped clouds in the bubble-trap experiments of Ref.~\cite{Carollo2022}, which can be larger than \(1~\mathrm{mm}\) with the shell thickness on the micrometer scale. Meanwhile, the proposed phase-separation generated shell structures can have cloud size ranging from \( 7.6\) to \(27.4~\mu\mathrm{m}\)~\cite{PhysRevA.106.013309}. The distinct length scales support the theoretical treatment of vortices as point-like topological objects on a spherical surface, an assumption to simplify the modeling of spherical vortex lattices. Although the spherical-shell structures available in experiments are large compared to a single vortex, realizing a huge quantity of vortices on a spherical surface still needs efforts to further increase the shell size and the strength of the effective monopole while shrinking the vortex cores.

We caution that in bosonic superfluids, the density is directly tied to the condensate order parameter, so a vortex appears as a hole in the density profile~\cite{FetterWalecka2003,Pethick-BEC}. By contrast, only the order parameter in BCS fermionic superfluids vanishes inside the vortex core but the particle density needs not vanish completely due to the presence of unpaired fermions\cite{PhysRevA.73.041603}, making vortices generally less visible in direct density measurements. Nevertheless, Ref.~\cite{Zwierlein2005VorticesAS} has shown that by an interaction quench to the strongly pairing side, the vortices of a BCS superfluid become visible as the density dips inside the vortex cores after the quench into the BEC regime. Similar techniques may be adapted to detect vortices of atomic Fermi superfluid on a spherical surface in the future.

On the theoretical side, an earlier study of spherical vortex systems based on elasticity theory predicted that the twelve fivefold disclinations required by Euler's theorem would be accompanied by additional 5–7 dislocation pairs (with one sevenfold site adjacent to a fivefold site) to recover the planar ground-state energy in the thermodynamic limit \cite{PhysRevB.55.3816}. Numerical studies on the Thomson problem found that such dislocation arrays indeed lower the energy in sufficiently large systems and showed visual evidence of the 5–7 dislocation pairs \cite{PhysRevB.56.3640}, but the Coulomb interaction of the Thomson problem is different from the nonlinear interaction of the GL theory. In contrast, our study of Fermi superfluid on a spherical surface takes different approaches by prescribing the vortex positions according to a a scaffold or obtaining the vortex structure by numerically minimizing the Abrikosov parameter $\beta_A$. As shown in Fig.~\ref{fig:extra}, the extrapolations of $\beta_A$ of the Fibonacci-lattice and minimization solutions in the large-$N_v$ limit both converge to $\beta_A \approx 1.16$, which corresponds to that of the triangular lattice of the planar case. However, we have not seen clear 5-7 dislocation pairs in both geometric and minimization constructions, as illustrated in Fig.~\ref{fig:compare}. Therefore, the spherical vortex systems of Fermi superfluids may be quasi-crystalline in nature without an obvious network of 5–7 dislocation pairs.


\section{Conclusion}\label{Sec:Conclusion}
Given the constraint that perfect lattice structures on a spherical surface with more than $20$ vortices are impossible, we have investigated approximate vortex-lattice structures of Fermi superfluid on a spherical surface given by different constructions according to the Ginzburg-Landau theory in the presence of a monopole field. While the geometric constructions using the random, geodesic-dome, and Fibonacci lattices as scaffolds provide intuitive structures based on the degenerate monopole harmonics, the minimization construction can generate configurations with lower nonlinear energies. Each vortex is identified as a zero of the order parameter and verified by the gauge-invariant current circulating around the vortex core. As the number of vortices increases, the Fibonacci lattice converges towards the minimization solution and provides an asymptotic approximation. The various possibilities of vortex configurations on a spherical surface thus illustrate interesting physics interfacing geometry, many-body systems, and collective excitations.

\begin{acknowledgments}
K. S. was supported by the NSF (Grant No. PHY-2310656). C. C. C. was supported by the DOE (Grant No. DE-SC0025809). We thank Dr. Shizeng Lin for stimulating discussions. Part of the research was conducted using Pinnacles (NSF MRI, No. 2019144) at the Cyberinfrastructure and Research Technologies (CIRT) at University of California, Merced.
\end{acknowledgments}

\appendix


\begin{thebibliography}{65}%
	\makeatletter
	\providecommand \@ifxundefined [1]{%
		\@ifx{#1\undefined}
	}%
	\providecommand \@ifnum [1]{%
		\ifnum #1\expandafter \@firstoftwo
		\else \expandafter \@secondoftwo
		\fi
	}%
	\providecommand \@ifx [1]{%
		\ifx #1\expandafter \@firstoftwo
		\else \expandafter \@secondoftwo
		\fi
	}%
	\providecommand \natexlab [1]{#1}%
	\providecommand \enquote  [1]{``#1''}%
	\providecommand \bibnamefont  [1]{#1}%
	\providecommand \bibfnamefont [1]{#1}%
	\providecommand \citenamefont [1]{#1}%
	\providecommand \href@noop [0]{\@secondoftwo}%
	\providecommand \href [0]{\begingroup \@sanitize@url \@href}%
	\providecommand \@href[1]{\@@startlink{#1}\@@href}%
	\providecommand \@@href[1]{\endgroup#1\@@endlink}%
	\providecommand \@sanitize@url [0]{\catcode `\\12\catcode `\$12\catcode
		`\&12\catcode `\#12\catcode `\^12\catcode `\_12\catcode `\%12\relax}%
	\providecommand \@@startlink[1]{}%
	\providecommand \@@endlink[0]{}%
	\providecommand \url  [0]{\begingroup\@sanitize@url \@url }%
	\providecommand \@url [1]{\endgroup\@href {#1}{\urlprefix }}%
	\providecommand \urlprefix  [0]{URL }%
	\providecommand \Eprint [0]{\href }%
	\providecommand \doibase [0]{https://doi.org/}%
	\providecommand \selectlanguage [0]{\@gobble}%
	\providecommand \bibinfo  [0]{\@secondoftwo}%
	\providecommand \bibfield  [0]{\@secondoftwo}%
	\providecommand \translation [1]{[#1]}%
	\providecommand \BibitemOpen [0]{}%
	\providecommand \bibitemStop [0]{}%
	\providecommand \bibitemNoStop [0]{.\EOS\space}%
	\providecommand \EOS [0]{\spacefactor3000\relax}%
	\providecommand \BibitemShut  [1]{\csname bibitem#1\endcsname}%
	\let\auto@bib@innerbib\@empty
	\bibitem [{\citenamefont {Carollo}\ \emph {et~al.}(2022)\citenamefont
		{Carollo}, \citenamefont {Aveline}, \citenamefont {Rhyno}, \citenamefont
		{Vishveshwara}, \citenamefont {Lannert}, \citenamefont {Murphree},
		\citenamefont {Elliott}, \citenamefont {Williams}, \citenamefont {Thompson},\
		and\ \citenamefont {Lundblad}}]{Carollo2022}%
	\BibitemOpen
	\bibfield  {author} {\bibinfo {author} {\bibfnamefont {R.~A.}\ \bibnamefont
			{Carollo}}, \bibinfo {author} {\bibfnamefont {D.~C.}\ \bibnamefont
			{Aveline}}, \bibinfo {author} {\bibfnamefont {B.}~\bibnamefont {Rhyno}},
		\bibinfo {author} {\bibfnamefont {S.}~\bibnamefont {Vishveshwara}}, \bibinfo
		{author} {\bibfnamefont {C.}~\bibnamefont {Lannert}}, \bibinfo {author}
		{\bibfnamefont {J.~D.}\ \bibnamefont {Murphree}}, \bibinfo {author}
		{\bibfnamefont {E.~R.}\ \bibnamefont {Elliott}}, \bibinfo {author}
		{\bibfnamefont {J.~R.}\ \bibnamefont {Williams}}, \bibinfo {author}
		{\bibfnamefont {R.~J.}\ \bibnamefont {Thompson}},\ and\ \bibinfo {author}
		{\bibfnamefont {N.}~\bibnamefont {Lundblad}},\ }\bibfield  {title} {\bibinfo
		{title} {Observation of ultracold atomic bubbles in orbital microgravity},\
	}\href {https://doi.org/10.1038/s41586-022-04639-8} {\bibfield  {journal}
		{\bibinfo  {journal} {Nature}\ }\textbf {\bibinfo {volume} {606}},\ \bibinfo
		{pages} {281} (\bibinfo {year} {2022})}\BibitemShut {NoStop}%
	\bibitem [{\citenamefont {Lundblad}\ \emph {et~al.}(2023)\citenamefont
		{Lundblad}, \citenamefont {Aveline}, \citenamefont {Balaž}, \citenamefont
		{Bentine}, \citenamefont {Bigelow}, \citenamefont {Boegel}, \citenamefont
		{Efremov}, \citenamefont {Gaaloul}, \citenamefont {Meister}, \citenamefont
		{Olshanii}, \citenamefont {Sá~de Melo}, \citenamefont {Tononi},
		\citenamefont {Vishveshwara}, \citenamefont {White}, \citenamefont {Wolf},\
		and\ \citenamefont {Garraway}}]{Lundblad_2023}%
	\BibitemOpen
	\bibfield  {author} {\bibinfo {author} {\bibfnamefont {N.}~\bibnamefont
			{Lundblad}}, \bibinfo {author} {\bibfnamefont {D.~C.}\ \bibnamefont
			{Aveline}}, \bibinfo {author} {\bibfnamefont {A.}~\bibnamefont {Balaž}},
		\bibinfo {author} {\bibfnamefont {E.}~\bibnamefont {Bentine}}, \bibinfo
		{author} {\bibfnamefont {N.~P.}\ \bibnamefont {Bigelow}}, \bibinfo {author}
		{\bibfnamefont {P.}~\bibnamefont {Boegel}}, \bibinfo {author} {\bibfnamefont
			{M.~A.}\ \bibnamefont {Efremov}}, \bibinfo {author} {\bibfnamefont
			{N.}~\bibnamefont {Gaaloul}}, \bibinfo {author} {\bibfnamefont
			{M.}~\bibnamefont {Meister}}, \bibinfo {author} {\bibfnamefont
			{M.}~\bibnamefont {Olshanii}}, \bibinfo {author} {\bibfnamefont {C.~A.~R.}\
			\bibnamefont {Sá~de Melo}}, \bibinfo {author} {\bibfnamefont
			{A.}~\bibnamefont {Tononi}}, \bibinfo {author} {\bibfnamefont
			{S.}~\bibnamefont {Vishveshwara}}, \bibinfo {author} {\bibfnamefont {A.~C.}\
			\bibnamefont {White}}, \bibinfo {author} {\bibfnamefont {A.}~\bibnamefont
			{Wolf}},\ and\ \bibinfo {author} {\bibfnamefont {B.~M.}\ \bibnamefont
			{Garraway}},\ }\bibfield  {title} {\bibinfo {title} {Perspective on quantum
			bubbles in microgravity},\ }\href {https://doi.org/10.1088/2058-9565/acb1cf}
	{\bibfield  {journal} {\bibinfo  {journal} {Quantum Sci. Technol.}\ }\textbf
		{\bibinfo {volume} {8}},\ \bibinfo {pages} {024003} (\bibinfo {year}
		{2023})}\BibitemShut {NoStop}%
	\bibitem [{\citenamefont {F\"olling}\ \emph {et~al.}(2006)\citenamefont
		{F\"olling}, \citenamefont {Widera}, \citenamefont {M\"uller}, \citenamefont
		{Gerbier},\ and\ \citenamefont {Bloch}}]{PhysRevLett.97.060403}%
	\BibitemOpen
	\bibfield  {author} {\bibinfo {author} {\bibfnamefont {S.}~\bibnamefont
			{F\"olling}}, \bibinfo {author} {\bibfnamefont {A.}~\bibnamefont {Widera}},
		\bibinfo {author} {\bibfnamefont {T.}~\bibnamefont {M\"uller}}, \bibinfo
		{author} {\bibfnamefont {F.}~\bibnamefont {Gerbier}},\ and\ \bibinfo {author}
		{\bibfnamefont {I.}~\bibnamefont {Bloch}},\ }\bibfield  {title} {\bibinfo
		{title} {Formation of spatial shell structure in the superfluid to mott
			insulator transition},\ }\href
	{https://doi.org/10.1103/PhysRevLett.97.060403} {\bibfield  {journal}
		{\bibinfo  {journal} {Phys. Rev. Lett.}\ }\textbf {\bibinfo {volume} {97}},\
		\bibinfo {pages} {060403} (\bibinfo {year} {2006})}\BibitemShut {NoStop}%
	\bibitem [{\citenamefont {Jia}\ \emph {et~al.}(2022)\citenamefont {Jia},
		\citenamefont {Huang}, \citenamefont {Qiu}, \citenamefont {Zhou},
		\citenamefont {Yan},\ and\ \citenamefont {Wang}}]{PhysRevLett.129.243402}%
	\BibitemOpen
	\bibfield  {author} {\bibinfo {author} {\bibfnamefont {F.}~\bibnamefont
			{Jia}}, \bibinfo {author} {\bibfnamefont {Z.}~\bibnamefont {Huang}}, \bibinfo
		{author} {\bibfnamefont {L.}~\bibnamefont {Qiu}}, \bibinfo {author}
		{\bibfnamefont {R.}~\bibnamefont {Zhou}}, \bibinfo {author} {\bibfnamefont
			{Y.}~\bibnamefont {Yan}},\ and\ \bibinfo {author} {\bibfnamefont
			{D.}~\bibnamefont {Wang}},\ }\bibfield  {title} {\bibinfo {title} {Expansion
			dynamics of a shell-shaped bose-einstein condensate},\ }\href
	{https://doi.org/10.1103/PhysRevLett.129.243402} {\bibfield  {journal}
		{\bibinfo  {journal} {Phys. Rev. Lett.}\ }\textbf {\bibinfo {volume} {129}},\
		\bibinfo {pages} {243402} (\bibinfo {year} {2022})}\BibitemShut {NoStop}%
	\bibitem [{\citenamefont {Bereta}\ \emph {et~al.}(2019)\citenamefont {Bereta},
		\citenamefont {Madeira}, \citenamefont {Bagnato},\ and\ \citenamefont
		{Caracanhas}}]{articlenew}%
	\BibitemOpen
	\bibfield  {author} {\bibinfo {author} {\bibfnamefont {S.}~\bibnamefont
			{Bereta}}, \bibinfo {author} {\bibfnamefont {L.}~\bibnamefont {Madeira}},
		\bibinfo {author} {\bibfnamefont {V.}~\bibnamefont {Bagnato}},\ and\ \bibinfo
		{author} {\bibfnamefont {M.}~\bibnamefont {Caracanhas}},\ }\bibfield  {title}
	{\bibinfo {title} {Bose–einstein condensation in spherically symmetric
			traps},\ }\href {https://doi.org/10.1119/1.5125092} {\bibfield  {journal}
		{\bibinfo  {journal} {Am. J. Phys.}\ }\textbf {\bibinfo {volume} {87}},\
		\bibinfo {pages} {924} (\bibinfo {year} {2019})}\BibitemShut {NoStop}%
	\bibitem [{\citenamefont {Tononi}\ and\ \citenamefont
		{Salasnich}(2023)}]{TononiReview23}%
	\BibitemOpen
	\bibfield  {author} {\bibinfo {author} {\bibfnamefont {A.}~\bibnamefont
			{Tononi}}\ and\ \bibinfo {author} {\bibfnamefont {L.}~\bibnamefont
			{Salasnich}},\ }\bibfield  {title} {\bibinfo {title} {Low-dimensional quantum
			gases in curved geometries},\ }\href
	{https://doi.org/10.1038/s42254-023-00591-2} {\bibfield  {journal} {\bibinfo
			{journal} {Nat. Rev. Phys.}\ }\textbf {\bibinfo {volume} {5}},\ \bibinfo
		{pages} {398} (\bibinfo {year} {2023})}\BibitemShut {NoStop}%
	\bibitem [{\citenamefont {Tononi}\ and\ \citenamefont
		{Salasnich}(2024)}]{TONONI20241}%
	\BibitemOpen
	\bibfield  {author} {\bibinfo {author} {\bibfnamefont {A.}~\bibnamefont
			{Tononi}}\ and\ \bibinfo {author} {\bibfnamefont {L.}~\bibnamefont
			{Salasnich}},\ }\bibfield  {title} {\bibinfo {title} {Shell-shaped atomic
			gases},\ }\href
	{https://doi.org/https://doi.org/10.1016/j.physrep.2024.04.004} {\bibfield
		{journal} {\bibinfo  {journal} {Phys. Rep.}\ }\textbf {\bibinfo {volume}
			{1072}},\ \bibinfo {pages} {1} (\bibinfo {year} {2024})}\BibitemShut
	{NoStop}%
	\bibitem [{\citenamefont {Turner}\ \emph {et~al.}(2010)\citenamefont {Turner},
		\citenamefont {Vitelli},\ and\ \citenamefont {Nelson}}]{RevModPhys.82.1301}%
	\BibitemOpen
	\bibfield  {author} {\bibinfo {author} {\bibfnamefont {A.~M.}\ \bibnamefont
			{Turner}}, \bibinfo {author} {\bibfnamefont {V.}~\bibnamefont {Vitelli}},\
		and\ \bibinfo {author} {\bibfnamefont {D.~R.}\ \bibnamefont {Nelson}},\
	}\bibfield  {title} {\bibinfo {title} {Vortices on curved surfaces},\ }\href
	{https://doi.org/10.1103/RevModPhys.82.1301} {\bibfield  {journal} {\bibinfo
			{journal} {Rev. Mod. Phys.}\ }\textbf {\bibinfo {volume} {82}},\ \bibinfo
		{pages} {1301} (\bibinfo {year} {2010})}\BibitemShut {NoStop}%
	\bibitem [{\citenamefont {Bereta}\ \emph {et~al.}(2021)\citenamefont {Bereta},
		\citenamefont {Caracanhas},\ and\ \citenamefont
		{Fetter}}]{PhysRevA.103.053306}%
	\BibitemOpen
	\bibfield  {author} {\bibinfo {author} {\bibfnamefont {S.~J.}\ \bibnamefont
			{Bereta}}, \bibinfo {author} {\bibfnamefont {M.~A.}\ \bibnamefont
			{Caracanhas}},\ and\ \bibinfo {author} {\bibfnamefont {A.~L.}\ \bibnamefont
			{Fetter}},\ }\bibfield  {title} {\bibinfo {title} {Superfluid vortex dynamics
			on a spherical film},\ }\href {https://doi.org/10.1103/PhysRevA.103.053306}
	{\bibfield  {journal} {\bibinfo  {journal} {Phys. Rev. A}\ }\textbf {\bibinfo
			{volume} {103}},\ \bibinfo {pages} {053306} (\bibinfo {year}
		{2021})}\BibitemShut {NoStop}%
	\bibitem [{\citenamefont {White}(2024)}]{PhysRevA.109.013301}%
	\BibitemOpen
	\bibfield  {author} {\bibinfo {author} {\bibfnamefont {A.~C.}\ \bibnamefont
			{White}},\ }\bibfield  {title} {\bibinfo {title} {Triangular vortex lattices
			and giant vortices in rotating bubble bose-einstein condensates},\ }\href
	{https://doi.org/10.1103/PhysRevA.109.013301} {\bibfield  {journal} {\bibinfo
			{journal} {Phys. Rev. A}\ }\textbf {\bibinfo {volume} {109}},\ \bibinfo
		{pages} {013301} (\bibinfo {year} {2024})}\BibitemShut {NoStop}%
	\bibitem [{\citenamefont {Chen}\ \emph {et~al.}(2025)\citenamefont {Chen},
		\citenamefont {Jiang}, \citenamefont {Bai}, \citenamefont {Yang},\ and\
		\citenamefont {Zheng}}]{PhysRevA.111.033322}%
	\BibitemOpen
	\bibfield  {author} {\bibinfo {author} {\bibfnamefont {X.-Y.}\ \bibnamefont
			{Chen}}, \bibinfo {author} {\bibfnamefont {L.}~\bibnamefont {Jiang}},
		\bibinfo {author} {\bibfnamefont {W.-K.}\ \bibnamefont {Bai}}, \bibinfo
		{author} {\bibfnamefont {T.}~\bibnamefont {Yang}},\ and\ \bibinfo {author}
		{\bibfnamefont {J.-H.}\ \bibnamefont {Zheng}},\ }\bibfield  {title} {\bibinfo
		{title} {Synthetic half-integer magnetic monopole and single-vortex dynamics
			in spherical bose-einstein condensates},\ }\href
	{https://doi.org/10.1103/PhysRevA.111.033322} {\bibfield  {journal} {\bibinfo
			{journal} {Phys. Rev. A}\ }\textbf {\bibinfo {volume} {111}},\ \bibinfo
		{pages} {033322} (\bibinfo {year} {2025})}\BibitemShut {NoStop}%
	\bibitem [{\citenamefont {Padavi\ifmmode~\acute{c}\else \'{c}\fi{}}\ \emph
		{et~al.}(2020)\citenamefont {Padavi\ifmmode~\acute{c}\else \'{c}\fi{}},
		\citenamefont {Sun}, \citenamefont {Lannert},\ and\ \citenamefont
		{Vishveshwara}}]{PhysRevA.102.043305}%
	\BibitemOpen
	\bibfield  {author} {\bibinfo {author} {\bibfnamefont {K.}~\bibnamefont
			{Padavi\ifmmode~\acute{c}\else \'{c}\fi{}}}, \bibinfo {author} {\bibfnamefont
			{K.}~\bibnamefont {Sun}}, \bibinfo {author} {\bibfnamefont {C.}~\bibnamefont
			{Lannert}},\ and\ \bibinfo {author} {\bibfnamefont {S.}~\bibnamefont
			{Vishveshwara}},\ }\bibfield  {title} {\bibinfo {title} {Vortex-antivortex
			physics in shell-shaped bose-einstein condensates},\ }\href
	{https://doi.org/10.1103/PhysRevA.102.043305} {\bibfield  {journal} {\bibinfo
			{journal} {Phys. Rev. A}\ }\textbf {\bibinfo {volume} {102}},\ \bibinfo
		{pages} {043305} (\bibinfo {year} {2020})}\BibitemShut {NoStop}%
	\bibitem [{\citenamefont {Adhikari}(2012)}]{PhysRevA.85.053631}%
	\BibitemOpen
	\bibfield  {author} {\bibinfo {author} {\bibfnamefont {S.~K.}\ \bibnamefont
			{Adhikari}},\ }\bibfield  {title} {\bibinfo {title} {Dipolar bose-einstein
			condensate in a ring or in a shell},\ }\href
	{https://doi.org/10.1103/PhysRevA.85.053631} {\bibfield  {journal} {\bibinfo
			{journal} {Phys. Rev. A}\ }\textbf {\bibinfo {volume} {85}},\ \bibinfo
		{pages} {053631} (\bibinfo {year} {2012})}\BibitemShut {NoStop}%
	\bibitem [{\citenamefont {Tononi}\ \emph {et~al.}(2024)\citenamefont {Tononi},
		\citenamefont {Salasnich},\ and\ \citenamefont
		{Yakimenko}}]{10.1116/5.0211426}%
	\BibitemOpen
	\bibfield  {author} {\bibinfo {author} {\bibfnamefont {A.}~\bibnamefont
			{Tononi}}, \bibinfo {author} {\bibfnamefont {L.}~\bibnamefont {Salasnich}},\
		and\ \bibinfo {author} {\bibfnamefont {A.}~\bibnamefont {Yakimenko}},\
	}\bibfield  {title} {\bibinfo {title} {Quantum vortices in curved
			geometries},\ }\href {https://doi.org/10.1116/5.0211426} {\bibfield
		{journal} {\bibinfo  {journal} {AVS Quantum Sci.}\ }\textbf {\bibinfo
			{volume} {6}},\ \bibinfo {pages} {030502} (\bibinfo {year}
		{2024})}\BibitemShut {NoStop}%
	\bibitem [{\citenamefont {Arazo}\ \emph {et~al.}(2021)\citenamefont {Arazo},
		\citenamefont {Mayol},\ and\ \citenamefont {Guilleumas}}]{Arazo_2021}%
	\BibitemOpen
	\bibfield  {author} {\bibinfo {author} {\bibfnamefont {M.}~\bibnamefont
			{Arazo}}, \bibinfo {author} {\bibfnamefont {R.}~\bibnamefont {Mayol}},\ and\
		\bibinfo {author} {\bibfnamefont {M.}~\bibnamefont {Guilleumas}},\ }\bibfield
	{title} {\bibinfo {title} {Shell-shaped condensates with gravitational sag:
			contact and dipolar interactions},\ }\href
	{https://doi.org/10.1088/1367-2630/ac37c9} {\bibfield  {journal} {\bibinfo
			{journal} {New J. Phys.}\ }\textbf {\bibinfo {volume} {23}},\ \bibinfo
		{pages} {113040} (\bibinfo {year} {2021})}\BibitemShut {NoStop}%
	\bibitem [{\citenamefont {He}\ \emph {et~al.}(2022)\citenamefont {He},
		\citenamefont {Guo},\ and\ \citenamefont {Chien}}]{He22Sphere}%
	\BibitemOpen
	\bibfield  {author} {\bibinfo {author} {\bibfnamefont {Y.}~\bibnamefont
			{He}}, \bibinfo {author} {\bibfnamefont {H.}~\bibnamefont {Guo}},\ and\
		\bibinfo {author} {\bibfnamefont {C.-C.}\ \bibnamefont {Chien}},\ }\bibfield
	{title} {\bibinfo {title} {Bcs-bec crossover of atomic fermi superfluid in a
			spherical bubble trap},\ }\href {https://doi.org/10.1103/PhysRevA.105.033324}
	{\bibfield  {journal} {\bibinfo  {journal} {Phys. Rev. A}\ }\textbf {\bibinfo
			{volume} {105}},\ \bibinfo {pages} {033324} (\bibinfo {year}
		{2022})}\BibitemShut {NoStop}%
	\bibitem [{\citenamefont {Gladilin}\ \emph {et~al.}(2008)\citenamefont
		{Gladilin}, \citenamefont {Tempere}, \citenamefont {Silvera}, \citenamefont
		{Devreese},\ and\ \citenamefont {Moshchalkov}}]{PhysRevB.77.024512}%
	\BibitemOpen
	\bibfield  {author} {\bibinfo {author} {\bibfnamefont {V.~N.}\ \bibnamefont
			{Gladilin}}, \bibinfo {author} {\bibfnamefont {J.}~\bibnamefont {Tempere}},
		\bibinfo {author} {\bibfnamefont {I.~F.}\ \bibnamefont {Silvera}}, \bibinfo
		{author} {\bibfnamefont {J.~T.}\ \bibnamefont {Devreese}},\ and\ \bibinfo
		{author} {\bibfnamefont {V.~V.}\ \bibnamefont {Moshchalkov}},\ }\bibfield
	{title} {\bibinfo {title} {Vortices on a superconducting nanoshell: Phase
			diagram and dynamics},\ }\href {https://doi.org/10.1103/PhysRevB.77.024512}
	{\bibfield  {journal} {\bibinfo  {journal} {Phys. Rev. B}\ }\textbf {\bibinfo
			{volume} {77}},\ \bibinfo {pages} {024512} (\bibinfo {year}
		{2008})}\BibitemShut {NoStop}%
	\bibitem [{\citenamefont {He}\ and\ \citenamefont {Chien}(2023)}]{He23Vortex}%
	\BibitemOpen
	\bibfield  {author} {\bibinfo {author} {\bibfnamefont {Y.}~\bibnamefont
			{He}}\ and\ \bibinfo {author} {\bibfnamefont {C.-C.}\ \bibnamefont {Chien}},\
	}\bibfield  {title} {\bibinfo {title} {Vortex structure and spectrum of an
			atomic fermi superfluid in a spherical bubble trap},\ }\href
	{https://doi.org/10.1103/PhysRevA.108.053303} {\bibfield  {journal} {\bibinfo
			{journal} {Phys. Rev. A}\ }\textbf {\bibinfo {volume} {108}},\ \bibinfo
		{pages} {053303} (\bibinfo {year} {2023})}\BibitemShut {NoStop}%
	\bibitem [{\citenamefont {Abrikosov}(1957)}]{Abrikosov1957JETP}%
	\BibitemOpen
	\bibfield  {author} {\bibinfo {author} {\bibfnamefont {A.~A.}\ \bibnamefont
			{Abrikosov}},\ }\bibfield  {title} {\bibinfo {title} {On the magnetic
			properties of superconductors of the second group},\ }\href@noop {}
	{\bibfield  {journal} {\bibinfo  {journal} {Sov. Phys. JETP}\ }\textbf
		{\bibinfo {volume} {5}},\ \bibinfo {pages} {1174} (\bibinfo {year}
		{1957})}\BibitemShut {NoStop}%
	\bibitem [{\citenamefont {Jones}\ and\ \citenamefont
		{March}(1985)}]{JonesMarch_TSSP2_Dover1985}%
	\BibitemOpen
	\bibfield  {author} {\bibinfo {author} {\bibfnamefont {W.}~\bibnamefont
			{Jones}}\ and\ \bibinfo {author} {\bibfnamefont {N.~H.}\ \bibnamefont
			{March}},\ }\href@noop {} {\emph {\bibinfo {title} {Theoretical Solid State
				Physics, Vol. 2: Non-Equilibrium and Disorder}}}\ (\bibinfo  {publisher}
	{Dover Publications},\ \bibinfo {address} {New York},\ \bibinfo {year}
	{1985})\BibitemShut {NoStop}%
	\bibitem [{\citenamefont {Kleiner}\ \emph {et~al.}(1964)\citenamefont
		{Kleiner}, \citenamefont {Roth},\ and\ \citenamefont
		{Autler}}]{PhysRev.133.A1226}%
	\BibitemOpen
	\bibfield  {author} {\bibinfo {author} {\bibfnamefont {W.~H.}\ \bibnamefont
			{Kleiner}}, \bibinfo {author} {\bibfnamefont {L.~M.}\ \bibnamefont {Roth}},\
		and\ \bibinfo {author} {\bibfnamefont {S.~H.}\ \bibnamefont {Autler}},\
	}\bibfield  {title} {\bibinfo {title} {Bulk solution of ginzburg-landau
			equations for type ii superconductors: Upper critical field region},\ }\href
	{https://doi.org/10.1103/PhysRev.133.A1226} {\bibfield  {journal} {\bibinfo
			{journal} {Phys. Rev.}\ }\textbf {\bibinfo {volume} {133}},\ \bibinfo {pages}
		{A1226} (\bibinfo {year} {1964})}\BibitemShut {NoStop}%
	\bibitem [{\citenamefont {Cromwell}(1997)}]{Cromwell1997}%
	\BibitemOpen
	\bibfield  {author} {\bibinfo {author} {\bibfnamefont {P.~R.}\ \bibnamefont
			{Cromwell}},\ }\href@noop {} {\emph {\bibinfo {title} {Polyhedra}}}\
	(\bibinfo  {publisher} {Cambridge University Press},\ \bibinfo {address}
	{Cambridge},\ \bibinfo {year} {1997})\BibitemShut {NoStop}%
	\bibitem [{\citenamefont {Altschuler}\ \emph {et~al.}(1997)\citenamefont
		{Altschuler}, \citenamefont {Williams}, \citenamefont {Ratner}, \citenamefont
		{Tipton}, \citenamefont {Stong}, \citenamefont {Dowla},\ and\ \citenamefont
		{Wooten}}]{PhysRevLett.78.2681}%
	\BibitemOpen
	\bibfield  {author} {\bibinfo {author} {\bibfnamefont {E.~L.}\ \bibnamefont
			{Altschuler}}, \bibinfo {author} {\bibfnamefont {T.~J.}\ \bibnamefont
			{Williams}}, \bibinfo {author} {\bibfnamefont {E.~R.}\ \bibnamefont
			{Ratner}}, \bibinfo {author} {\bibfnamefont {R.}~\bibnamefont {Tipton}},
		\bibinfo {author} {\bibfnamefont {R.}~\bibnamefont {Stong}}, \bibinfo
		{author} {\bibfnamefont {F.}~\bibnamefont {Dowla}},\ and\ \bibinfo {author}
		{\bibfnamefont {F.}~\bibnamefont {Wooten}},\ }\bibfield  {title} {\bibinfo
		{title} {Possible global minimum lattice configurations for thomson's problem
			of charges on a sphere},\ }\href
	{https://doi.org/10.1103/PhysRevLett.78.2681} {\bibfield  {journal} {\bibinfo
			{journal} {Phys. Rev. Lett.}\ }\textbf {\bibinfo {volume} {78}},\ \bibinfo
		{pages} {2681} (\bibinfo {year} {1997})}\BibitemShut {NoStop}%
	\bibitem [{\citenamefont {Bowick}\ \emph {et~al.}(2002)\citenamefont {Bowick},
		\citenamefont {Cacciuto}, \citenamefont {Nelson},\ and\ \citenamefont
		{Travesset}}]{PhysRevLett.89.185502}%
	\BibitemOpen
	\bibfield  {author} {\bibinfo {author} {\bibfnamefont {M.}~\bibnamefont
			{Bowick}}, \bibinfo {author} {\bibfnamefont {A.}~\bibnamefont {Cacciuto}},
		\bibinfo {author} {\bibfnamefont {D.~R.}\ \bibnamefont {Nelson}},\ and\
		\bibinfo {author} {\bibfnamefont {A.}~\bibnamefont {Travesset}},\ }\bibfield
	{title} {\bibinfo {title} {Crystalline order on a sphere and the generalized
			thomson problem},\ }\href {https://doi.org/10.1103/PhysRevLett.89.185502}
	{\bibfield  {journal} {\bibinfo  {journal} {Phys. Rev. Lett.}\ }\textbf
		{\bibinfo {volume} {89}},\ \bibinfo {pages} {185502} (\bibinfo {year}
		{2002})}\BibitemShut {NoStop}%
	\bibitem [{\citenamefont {Bausch}\ \emph {et~al.}(2003)\citenamefont {Bausch},
		\citenamefont {Bowick}, \citenamefont {Cacciuto}, \citenamefont {Dinsmore},
		\citenamefont {Hsu}, \citenamefont {Nelson}, \citenamefont {Nikolaides},\
		and\ \citenamefont {Weitz}}]{article6}%
	\BibitemOpen
	\bibfield  {author} {\bibinfo {author} {\bibfnamefont {A.}~\bibnamefont
			{Bausch}}, \bibinfo {author} {\bibfnamefont {M.}~\bibnamefont {Bowick}},
		\bibinfo {author} {\bibfnamefont {A.}~\bibnamefont {Cacciuto}}, \bibinfo
		{author} {\bibfnamefont {A.~D.}\ \bibnamefont {Dinsmore}}, \bibinfo {author}
		{\bibfnamefont {M.}~\bibnamefont {Hsu}}, \bibinfo {author} {\bibfnamefont
			{D.}~\bibnamefont {Nelson}}, \bibinfo {author} {\bibfnamefont
			{M.}~\bibnamefont {Nikolaides}},\ and\ \bibinfo {author} {\bibfnamefont
			{D.~A.}\ \bibnamefont {Weitz}},\ }\bibfield  {title} {\bibinfo {title} {Grain
			boundary scars and spherical crystallography},\ }\href
	{https://doi.org/10.1126/science.1081160} {\bibfield  {journal} {\bibinfo
			{journal} {Science}\ }\textbf {\bibinfo {volume} {299}},\ \bibinfo {pages}
		{1716} (\bibinfo {year} {2003})}\BibitemShut {NoStop}%
	\bibitem [{\citenamefont {Haldane}(1983)}]{PhysRevLett.51.605}%
	\BibitemOpen
	\bibfield  {author} {\bibinfo {author} {\bibfnamefont {F.~D.~M.}\
			\bibnamefont {Haldane}},\ }\bibfield  {title} {\bibinfo {title} {Fractional
			quantization of the hall effect: A hierarchy of incompressible quantum fluid
			states},\ }\href {https://doi.org/10.1103/PhysRevLett.51.605} {\bibfield
		{journal} {\bibinfo  {journal} {Phys. Rev. Lett.}\ }\textbf {\bibinfo
			{volume} {51}},\ \bibinfo {pages} {605} (\bibinfo {year} {1983})}\BibitemShut
	{NoStop}%
	\bibitem [{\citenamefont {Kvorning}(2013)}]{PhysRevB.87.195131}%
	\BibitemOpen
	\bibfield  {author} {\bibinfo {author} {\bibfnamefont {T.}~\bibnamefont
			{Kvorning}},\ }\bibfield  {title} {\bibinfo {title} {Quantum hall hierarchy
			in a spherical geometry},\ }\href
	{https://doi.org/10.1103/PhysRevB.87.195131} {\bibfield  {journal} {\bibinfo
			{journal} {Phys. Rev. B}\ }\textbf {\bibinfo {volume} {87}},\ \bibinfo
		{pages} {195131} (\bibinfo {year} {2013})}\BibitemShut {NoStop}%
	\bibitem [{\citenamefont {Fetter}\ and\ \citenamefont
		{Walecka}(2003)}]{FetterWalecka2003}%
	\BibitemOpen
	\bibfield  {author} {\bibinfo {author} {\bibfnamefont {A.~L.}\ \bibnamefont
			{Fetter}}\ and\ \bibinfo {author} {\bibfnamefont {J.~D.}\ \bibnamefont
			{Walecka}},\ }\href@noop {} {\emph {\bibinfo {title} {Quantum Theory of
				Many-Particle Systems}}}\ (\bibinfo  {publisher} {Dover Publications},\
	\bibinfo {address} {Mineola, NY},\ \bibinfo {year} {2003})\ \bibinfo {note}
	{unabridged republication of the 1971 edition}\BibitemShut {NoStop}%
	\bibitem [{\citenamefont {Mazenko}(2003)}]{Mazenko2003FOD}%
	\BibitemOpen
	\bibfield  {author} {\bibinfo {author} {\bibfnamefont {G.~F.}\ \bibnamefont
			{Mazenko}},\ }\href {https://doi.org/10.1002/9783527618958} {\emph {\bibinfo
			{title} {Fluctuations, Order, and Defects}}}\ (\bibinfo  {publisher}
	{Wiley-VCH},\ \bibinfo {address} {Weinheim, Germany},\ \bibinfo {year}
	{2003})\BibitemShut {NoStop}%
	\bibitem [{\citenamefont {Dirac}(1931)}]{Dirac1931}%
	\BibitemOpen
	\bibfield  {author} {\bibinfo {author} {\bibfnamefont {P.~A.~M.}\
			\bibnamefont {Dirac}},\ }\bibfield  {title} {\bibinfo {title} {Quantised
			singularities in the electromagnetic field,},\ }\href
	{https://doi.org/10.1098/rspa.1931.0130} {\bibfield  {journal} {\bibinfo
			{journal} {Proc. R. Soc. A}\ }\textbf {\bibinfo {volume} {133}},\ \bibinfo
		{pages} {60} (\bibinfo {year} {1931})},\ \Eprint
	{https://arxiv.org/abs/https://royalsocietypublishing.org/rspa/article-pdf/133/821/60/26949/rspa.1931.0130.pdf}
	{https://royalsocietypublishing.org/rspa/article-pdf/133/821/60/26949/rspa.1931.0130.pdf}
	\BibitemShut {NoStop}%
	\bibitem [{\citenamefont {Pietil\"a}\ and\ \citenamefont
		{M\"ott\"onen}(2009)}]{PhysRevLett.103.030401}%
	\BibitemOpen
	\bibfield  {author} {\bibinfo {author} {\bibfnamefont {V.}~\bibnamefont
			{Pietil\"a}}\ and\ \bibinfo {author} {\bibfnamefont {M.}~\bibnamefont
			{M\"ott\"onen}},\ }\bibfield  {title} {\bibinfo {title} {Creation of dirac
			monopoles in spinor bose-einstein condensates},\ }\href
	{https://doi.org/10.1103/PhysRevLett.103.030401} {\bibfield  {journal}
		{\bibinfo  {journal} {Phys. Rev. Lett.}\ }\textbf {\bibinfo {volume} {103}},\
		\bibinfo {pages} {030401} (\bibinfo {year} {2009})}\BibitemShut {NoStop}%
	\bibitem [{\citenamefont {Zhou}\ \emph {et~al.}(2018)\citenamefont {Zhou},
		\citenamefont {Wu}, \citenamefont {Guo}, \citenamefont {Wang}, \citenamefont
		{Pu},\ and\ \citenamefont {Zhou}}]{PhysRevLett.120.130402}%
	\BibitemOpen
	\bibfield  {author} {\bibinfo {author} {\bibfnamefont {X.-F.}\ \bibnamefont
			{Zhou}}, \bibinfo {author} {\bibfnamefont {C.}~\bibnamefont {Wu}}, \bibinfo
		{author} {\bibfnamefont {G.-C.}\ \bibnamefont {Guo}}, \bibinfo {author}
		{\bibfnamefont {R.}~\bibnamefont {Wang}}, \bibinfo {author} {\bibfnamefont
			{H.}~\bibnamefont {Pu}},\ and\ \bibinfo {author} {\bibfnamefont {Z.-W.}\
			\bibnamefont {Zhou}},\ }\bibfield  {title} {\bibinfo {title} {Synthetic
			landau levels and spinor vortex matter on a haldane spherical surface with a
			magnetic monopole},\ }\href {https://doi.org/10.1103/PhysRevLett.120.130402}
	{\bibfield  {journal} {\bibinfo  {journal} {Phys. Rev. Lett.}\ }\textbf
		{\bibinfo {volume} {120}},\ \bibinfo {pages} {130402} (\bibinfo {year}
		{2018})}\BibitemShut {NoStop}%
	\bibitem [{\citenamefont {Ray}\ \emph {et~al.}(2014)\citenamefont {Ray},
		\citenamefont {Ruokokoski}, \citenamefont {Kandel}, \citenamefont
		{Möttönen},\ and\ \citenamefont {Hall}}]{Ray_2014}%
	\BibitemOpen
	\bibfield  {author} {\bibinfo {author} {\bibfnamefont {M.~W.}\ \bibnamefont
			{Ray}}, \bibinfo {author} {\bibfnamefont {E.}~\bibnamefont {Ruokokoski}},
		\bibinfo {author} {\bibfnamefont {S.}~\bibnamefont {Kandel}}, \bibinfo
		{author} {\bibfnamefont {M.}~\bibnamefont {Möttönen}},\ and\ \bibinfo
		{author} {\bibfnamefont {D.~S.}\ \bibnamefont {Hall}},\ }\bibfield  {title}
	{\bibinfo {title} {Observation of dirac monopoles in a synthetic magnetic
			field},\ }\href {https://doi.org/10.1038/nature12954} {\bibfield  {journal}
		{\bibinfo  {journal} {Nature}\ }\textbf {\bibinfo {volume} {505}},\ \bibinfo
		{pages} {657–660} (\bibinfo {year} {2014})}\BibitemShut {NoStop}%
	\bibitem [{\citenamefont {Ollikainen}\ \emph {et~al.}(2017)\citenamefont
		{Ollikainen}, \citenamefont {Tiurev}, \citenamefont {Blinova}, \citenamefont
		{Lee}, \citenamefont {Hall},\ and\ \citenamefont
		{M\"ott\"onen}}]{PhysRevX.7.021023}%
	\BibitemOpen
	\bibfield  {author} {\bibinfo {author} {\bibfnamefont {T.}~\bibnamefont
			{Ollikainen}}, \bibinfo {author} {\bibfnamefont {K.}~\bibnamefont {Tiurev}},
		\bibinfo {author} {\bibfnamefont {A.}~\bibnamefont {Blinova}}, \bibinfo
		{author} {\bibfnamefont {W.}~\bibnamefont {Lee}}, \bibinfo {author}
		{\bibfnamefont {D.~S.}\ \bibnamefont {Hall}},\ and\ \bibinfo {author}
		{\bibfnamefont {M.}~\bibnamefont {M\"ott\"onen}},\ }\bibfield  {title}
	{\bibinfo {title} {Experimental realization of a dirac monopole through the
			decay of an isolated monopole},\ }\href
	{https://doi.org/10.1103/PhysRevX.7.021023} {\bibfield  {journal} {\bibinfo
			{journal} {Phys. Rev. X}\ }\textbf {\bibinfo {volume} {7}},\ \bibinfo {pages}
		{021023} (\bibinfo {year} {2017})}\BibitemShut {NoStop}%
	\bibitem [{\citenamefont {Lima}\ \emph {et~al.}(2018)\citenamefont {Lima},
		\citenamefont {Santos}, \citenamefont {Cunha},\ and\ \citenamefont
		{Moraes}}]{article3}%
	\BibitemOpen
	\bibfield  {author} {\bibinfo {author} {\bibfnamefont {J.}~\bibnamefont
			{Lima}}, \bibinfo {author} {\bibfnamefont {A.}~\bibnamefont {Santos}},
		\bibinfo {author} {\bibfnamefont {M.}~\bibnamefont {Cunha}},\ and\ \bibinfo
		{author} {\bibfnamefont {F.}~\bibnamefont {Moraes}},\ }\bibfield  {title}
	{\bibinfo {title} {Effects of rotation on landau states of electrons on a
			spherical shell},\ }\href {https://doi.org/10.1016/j.physleta.2018.07.029}
	{\bibfield  {journal} {\bibinfo  {journal} {Phys. Lett. A}\ }\textbf
		{\bibinfo {volume} {382}} (\bibinfo {year} {2018})}\BibitemShut {NoStop}%
	\bibitem [{\citenamefont {Du}\ and\ \citenamefont {Ju}(2005)}]{Du05}%
	\BibitemOpen
	\bibfield  {author} {\bibinfo {author} {\bibfnamefont {Q.}~\bibnamefont
			{Du}}\ and\ \bibinfo {author} {\bibfnamefont {L.}~\bibnamefont {Ju}},\
	}\bibfield  {title} {\bibinfo {title} {Approximations of a ginzburg-landau
			model for superconducting hollow spheres based on spherical centroidal
			voronoi tessellations},\ }\href
	{https://doi.org/10.1090/S0025-5718-04-01719-3} {\bibfield  {journal}
		{\bibinfo  {journal} {Math. Comp.}\ }\textbf {\bibinfo {volume} {74}},\
		\bibinfo {pages} {1257} (\bibinfo {year} {2005})}\BibitemShut {NoStop}%
	\bibitem [{\citenamefont {Simonucci}\ \emph {et~al.}(2015)\citenamefont
		{Simonucci}, \citenamefont {Pieri},\ and\ \citenamefont
		{Strinati}}]{Simonucci2015}%
	\BibitemOpen
	\bibfield  {author} {\bibinfo {author} {\bibfnamefont {S.}~\bibnamefont
			{Simonucci}}, \bibinfo {author} {\bibfnamefont {P.}~\bibnamefont {Pieri}},\
		and\ \bibinfo {author} {\bibfnamefont {G.~C.}\ \bibnamefont {Strinati}},\
	}\bibfield  {title} {\bibinfo {title} {Vortex arrays in neutral trapped fermi
			gases through the bcs--bec crossover},\ }\href
	{https://doi.org/10.1038/nphys3449} {\bibfield  {journal} {\bibinfo
			{journal} {Nature Physics}\ }\textbf {\bibinfo {volume} {11}},\ \bibinfo
		{pages} {941} (\bibinfo {year} {2015})}\BibitemShut {NoStop}%
	\bibitem [{\citenamefont {Wu}\ and\ \citenamefont {Yang}(1976)}]{Wu1976}%
	\BibitemOpen
	\bibfield  {author} {\bibinfo {author} {\bibfnamefont {T.~T.}\ \bibnamefont
			{Wu}}\ and\ \bibinfo {author} {\bibfnamefont {C.~N.}\ \bibnamefont {Yang}},\
	}\bibfield  {title} {\bibinfo {title} {Dirac monopole without strings:
			Monopole harmonics},\ }\href {https://doi.org/10.1016/0550-3213(76)90143-7}
	{\bibfield  {journal} {\bibinfo  {journal} {Nucl. Phys. B}\ }\textbf
		{\bibinfo {volume} {107}},\ \bibinfo {pages} {365} (\bibinfo {year}
		{1976})}\BibitemShut {NoStop}%
	\bibitem [{\citenamefont {{Dodgson}}(1996)}]{1996JPhA...29.2499D}%
	\BibitemOpen
	\bibfield  {author} {\bibinfo {author} {\bibfnamefont {M.~J.~W.}\
			\bibnamefont {{Dodgson}}},\ }\bibfield  {title} {\bibinfo {title}
		{{Investigation on the ground states of a model thin-film superconductor on a
				sphere}},\ }\href {https://doi.org/10.1088/0305-4470/29/10/028} {\bibfield
		{journal} {\bibinfo  {journal} {J. Phys. A: Math. Gen.}\ }\textbf {\bibinfo
			{volume} {29}},\ \bibinfo {pages} {2499} (\bibinfo {year} {1996})},\ \Eprint
	{https://arxiv.org/abs/cond-mat/9512124} {arXiv:cond-mat/9512124 [cond-mat]}
	\BibitemShut {NoStop}%
	\bibitem [{\citenamefont {Dodgson}\ and\ \citenamefont
		{Moore}(1997)}]{PhysRevB.55.3816}%
	\BibitemOpen
	\bibfield  {author} {\bibinfo {author} {\bibfnamefont {M.~J.~W.}\
			\bibnamefont {Dodgson}}\ and\ \bibinfo {author} {\bibfnamefont {M.~A.}\
			\bibnamefont {Moore}},\ }\bibfield  {title} {\bibinfo {title} {Vortices in a
			thin-film superconductor with a spherical geometry},\ }\href
	{https://doi.org/10.1103/PhysRevB.55.3816} {\bibfield  {journal} {\bibinfo
			{journal} {Phys. Rev. B}\ }\textbf {\bibinfo {volume} {55}},\ \bibinfo
		{pages} {3816} (\bibinfo {year} {1997})}\BibitemShut {NoStop}%
	\bibitem [{\citenamefont {Hopf}(1927)}]{Hopf1927}%
	\BibitemOpen
	\bibfield  {author} {\bibinfo {author} {\bibfnamefont {H.}~\bibnamefont
			{Hopf}},\ }\bibfield  {title} {\bibinfo {title} {{Vektorfelder in
				n-dimensionalen Mannigfaltigkeiten}},\ }\href@noop {} {\bibfield  {journal}
		{\bibinfo  {journal} {Math. Ann.}\ }\textbf {\bibinfo {volume} {96}},\
		\bibinfo {pages} {225} (\bibinfo {year} {1927})}\BibitemShut {NoStop}%
	\bibitem [{\citenamefont {Milnor}(1997)}]{MilnorTopology}%
	\BibitemOpen
	\bibfield  {author} {\bibinfo {author} {\bibfnamefont {J.~W.}\ \bibnamefont
			{Milnor}},\ }\href@noop {} {\emph {\bibinfo {title} {Topology from the
				Differentiable Viewpoint}}}\ (\bibinfo  {publisher} {Princeton University
		Press},\ \bibinfo {address} {Princeton, NJ},\ \bibinfo {year}
	{1997})\BibitemShut {NoStop}%
	\bibitem [{\citenamefont {Goldman}\ \emph {et~al.}(2013)\citenamefont
		{Goldman}, \citenamefont {Juzeliūnas}, \citenamefont {{\"O}hberg},\ and\
		\citenamefont {Spielman}}]{Goldman2013LightinducedGF}%
	\BibitemOpen
	\bibfield  {author} {\bibinfo {author} {\bibfnamefont {N.}~\bibnamefont
			{Goldman}}, \bibinfo {author} {\bibfnamefont {G.}~\bibnamefont
			{Juzeliūnas}}, \bibinfo {author} {\bibfnamefont {P.}~\bibnamefont
			{{\"O}hberg}},\ and\ \bibinfo {author} {\bibfnamefont {I.~B.}\ \bibnamefont
			{Spielman}},\ }\bibfield  {title} {\bibinfo {title} {Light-induced gauge
			fields for ultracold atoms},\ }\href
	{https://api.semanticscholar.org/CorpusID:2829048} {\bibfield  {journal}
		{\bibinfo  {journal} {Rep. Prog. Phys.}\ }\textbf {\bibinfo {volume} {77}}
		(\bibinfo {year} {2013})}\BibitemShut {NoStop}%
	\bibitem [{\citenamefont {Dalibard}\ \emph {et~al.}(2011)\citenamefont
		{Dalibard}, \citenamefont {Gerbier}, \citenamefont
		{Juzeli\ifmmode~\bar{u}\else \={u}\fi{}nas},\ and\ \citenamefont
		{\"Ohberg}}]{RevModPhys.83.1523}%
	\BibitemOpen
	\bibfield  {author} {\bibinfo {author} {\bibfnamefont {J.}~\bibnamefont
			{Dalibard}}, \bibinfo {author} {\bibfnamefont {F.}~\bibnamefont {Gerbier}},
		\bibinfo {author} {\bibfnamefont {G.}~\bibnamefont
			{Juzeli\ifmmode~\bar{u}\else \={u}\fi{}nas}},\ and\ \bibinfo {author}
		{\bibfnamefont {P.}~\bibnamefont {\"Ohberg}},\ }\bibfield  {title} {\bibinfo
		{title} {Colloquium: Artificial gauge potentials for neutral atoms},\ }\href
	{https://doi.org/10.1103/RevModPhys.83.1523} {\bibfield  {journal} {\bibinfo
			{journal} {Rev. Mod. Phys.}\ }\textbf {\bibinfo {volume} {83}},\ \bibinfo
		{pages} {1523} (\bibinfo {year} {2011})}\BibitemShut {NoStop}%
	\bibitem [{\citenamefont {Lin}\ \emph {et~al.}(2009)\citenamefont {Lin},
		\citenamefont {Compton}, \citenamefont {Jimenez}, \citenamefont {Porto},\
		and\ \citenamefont {Spielman}}]{articleultracold}%
	\BibitemOpen
	\bibfield  {author} {\bibinfo {author} {\bibfnamefont {Y.-J.}\ \bibnamefont
			{Lin}}, \bibinfo {author} {\bibfnamefont {R.}~\bibnamefont {Compton}},
		\bibinfo {author} {\bibfnamefont {K.}~\bibnamefont {Jimenez}}, \bibinfo
		{author} {\bibfnamefont {J.}~\bibnamefont {Porto}},\ and\ \bibinfo {author}
		{\bibfnamefont {I.}~\bibnamefont {Spielman}},\ }\bibfield  {title} {\bibinfo
		{title} {Synthetic magnetic fields for ultracold neutral atoms},\ }\href
	{https://doi.org/10.1038/nature08609} {\bibfield  {journal} {\bibinfo
			{journal} {Nature}\ }\textbf {\bibinfo {volume} {462}},\ \bibinfo {pages}
		{628} (\bibinfo {year} {2009})}\BibitemShut {NoStop}%
	\bibitem [{\citenamefont {Tinkham}(1996)}]{Tinkham1996}%
	\BibitemOpen
	\bibfield  {author} {\bibinfo {author} {\bibfnamefont {M.}~\bibnamefont
			{Tinkham}},\ }\href@noop {} {\emph {\bibinfo {title} {Introduction to
				Superconductivity}}},\ \bibinfo {edition} {2nd}\ ed.\ (\bibinfo  {publisher}
	{McGraw--Hill},\ \bibinfo {address} {New York},\ \bibinfo {year}
	{1996})\BibitemShut {NoStop}%
	\bibitem [{\citenamefont {Campi}\ \emph {et~al.}(2025)\citenamefont {Campi},
		\citenamefont {Alimenti}, \citenamefont {Logvenov}, \citenamefont {Smith},
		\citenamefont {Balakirev}, \citenamefont {Lee}, \citenamefont {Balicas},
		\citenamefont {Silva}, \citenamefont {Ummarino}, \citenamefont {Midei},
		\citenamefont {Perali}, \citenamefont {Valletta},\ and\ \citenamefont
		{Bianconi}}]{k2yd-vpbn}%
	\BibitemOpen
	\bibfield  {author} {\bibinfo {author} {\bibfnamefont {G.}~\bibnamefont
			{Campi}}, \bibinfo {author} {\bibfnamefont {A.}~\bibnamefont {Alimenti}},
		\bibinfo {author} {\bibfnamefont {G.}~\bibnamefont {Logvenov}}, \bibinfo
		{author} {\bibfnamefont {G.~A.}\ \bibnamefont {Smith}}, \bibinfo {author}
		{\bibfnamefont {F.}~\bibnamefont {Balakirev}}, \bibinfo {author}
		{\bibfnamefont {S.-E.}\ \bibnamefont {Lee}}, \bibinfo {author} {\bibfnamefont
			{L.}~\bibnamefont {Balicas}}, \bibinfo {author} {\bibfnamefont
			{E.}~\bibnamefont {Silva}}, \bibinfo {author} {\bibfnamefont {G.~A.}\
			\bibnamefont {Ummarino}}, \bibinfo {author} {\bibfnamefont {G.}~\bibnamefont
			{Midei}}, \bibinfo {author} {\bibfnamefont {A.}~\bibnamefont {Perali}},
		\bibinfo {author} {\bibfnamefont {A.}~\bibnamefont {Valletta}},\ and\
		\bibinfo {author} {\bibfnamefont {A.}~\bibnamefont {Bianconi}},\ }\bibfield
	{title} {\bibinfo {title} {Upper critical magnetic field and multiband
			superconductivity in artificial $\text{high-}{T}_{c}$ superlattices of nano
			quantum wells},\ }\href {https://doi.org/10.1103/k2yd-vpbn} {\bibfield
		{journal} {\bibinfo  {journal} {Phys. Rev. Mater.}\ }\textbf {\bibinfo
			{volume} {9}},\ \bibinfo {pages} {074204} (\bibinfo {year}
		{2025})}\BibitemShut {NoStop}%
	\bibitem [{\citenamefont {Tamm}(1931)}]{Tamm1931}%
	\BibitemOpen
	\bibfield  {author} {\bibinfo {author} {\bibfnamefont {I.}~\bibnamefont
			{Tamm}},\ }\bibfield  {title} {\bibinfo {title} {Die verallgemeinerten
			kugelfunktionen und die wellenfunktionen eines elektrons im felde eines
			magnetpoles [the generalized spherical harmonics and the wave functions of an
			electron in the field of a magnetic pole]},\ }\href@noop {} {\bibfield
		{journal} {\bibinfo  {journal} {Z. Phys.}\ }\textbf {\bibinfo {volume}
			{71}},\ \bibinfo {pages} {141} (\bibinfo {year} {1931})}\BibitemShut
	{NoStop}%
	\bibitem [{\citenamefont {Sakurai}\ and\ \citenamefont
		{Napolitano}(2010)}]{MQM}%
	\BibitemOpen
	\bibfield  {author} {\bibinfo {author} {\bibfnamefont {J.~J.}\ \bibnamefont
			{Sakurai}}\ and\ \bibinfo {author} {\bibfnamefont {J.~J.}\ \bibnamefont
			{Napolitano}},\ }\href@noop {} {\emph {\bibinfo {title} {Modern quantum
				mechanics}}},\ \bibinfo {edition} {2nd}\ ed.\ (\bibinfo  {publisher}
	{Pearson},\ \bibinfo {address} {London, UK},\ \bibinfo {year}
	{2010})\BibitemShut {NoStop}%
	\bibitem [{\citenamefont {Tashiro}(1977)}]{tashiro1977methods}%
	\BibitemOpen
	\bibfield  {author} {\bibinfo {author} {\bibfnamefont {Y.}~\bibnamefont
			{Tashiro}},\ }\bibfield  {title} {\bibinfo {title} {On methods for generating
			uniform random points on the surface of a sphere},\ }\href
	{https://doi.org/10.1007/BF02532791} {\bibfield  {journal} {\bibinfo
			{journal} {Ann. Inst. Stat. Math.}\ }\textbf {\bibinfo {volume} {29}},\
		\bibinfo {pages} {295} (\bibinfo {year} {1977})}\BibitemShut {NoStop}%
	\bibitem [{\citenamefont
		{Wenninger}(2012)}]{WenningerSphericalModelsDover2012}%
	\BibitemOpen
	\bibfield  {author} {\bibinfo {author} {\bibfnamefont {M.~J.}\ \bibnamefont
			{Wenninger}},\ }\href@noop {} {\emph {\bibinfo {title} {Spherical Models}}}\
	(\bibinfo  {publisher} {Dover Publications},\ \bibinfo {address} {Mineola,
		NY},\ \bibinfo {year} {2012})\BibitemShut {NoStop}%
	\bibitem [{\citenamefont {Marques}\ \emph {et~al.}(2013)\citenamefont
		{Marques}, \citenamefont {Bouville}, \citenamefont {Ribardière},
		\citenamefont {Santos},\ and\ \citenamefont {Bouatouch}}]{Santos2013}%
	\BibitemOpen
	\bibfield  {author} {\bibinfo {author} {\bibfnamefont {R.}~\bibnamefont
			{Marques}}, \bibinfo {author} {\bibfnamefont {C.}~\bibnamefont {Bouville}},
		\bibinfo {author} {\bibfnamefont {M.}~\bibnamefont {Ribardière}}, \bibinfo
		{author} {\bibfnamefont {L.~P.}\ \bibnamefont {Santos}},\ and\ \bibinfo
		{author} {\bibfnamefont {K.}~\bibnamefont {Bouatouch}},\ }\bibfield  {title}
	{\bibinfo {title} {Spherical fibonacci point sets for illumination
			integrals},\ }\href {https://doi.org/https://doi.org/10.1111/cgf.12190}
	{\bibfield  {journal} {\bibinfo  {journal} {Computer Graphics Forum}\
		}\textbf {\bibinfo {volume} {32}},\ \bibinfo {pages} {134} (\bibinfo {year}
		{2013})},\ \Eprint
	{https://arxiv.org/abs/https://onlinelibrary.wiley.com/doi/pdf/10.1111/cgf.12190}
	{https://onlinelibrary.wiley.com/doi/pdf/10.1111/cgf.12190} \BibitemShut
	{NoStop}%
	\bibitem [{\citenamefont {Song}\ \emph {et~al.}(2022)\citenamefont {Song},
		\citenamefont {Gao}, \citenamefont {Hou}, \citenamefont {Wang}, \citenamefont
		{Zhou}, \citenamefont {He}, \citenamefont {Guo},\ and\ \citenamefont
		{Chien}}]{PhysRevResearch.4.023005}%
	\BibitemOpen
	\bibfield  {author} {\bibinfo {author} {\bibfnamefont {C.-H.}\ \bibnamefont
			{Song}}, \bibinfo {author} {\bibfnamefont {Q.-C.}\ \bibnamefont {Gao}},
		\bibinfo {author} {\bibfnamefont {X.-Y.}\ \bibnamefont {Hou}}, \bibinfo
		{author} {\bibfnamefont {X.}~\bibnamefont {Wang}}, \bibinfo {author}
		{\bibfnamefont {Z.}~\bibnamefont {Zhou}}, \bibinfo {author} {\bibfnamefont
			{Y.}~\bibnamefont {He}}, \bibinfo {author} {\bibfnamefont {H.}~\bibnamefont
			{Guo}},\ and\ \bibinfo {author} {\bibfnamefont {C.-C.}\ \bibnamefont
			{Chien}},\ }\bibfield  {title} {\bibinfo {title} {Machine learning of the
			$xy$ model on a spherical fibonacci lattice},\ }\href
	{https://doi.org/10.1103/PhysRevResearch.4.023005} {\bibfield  {journal}
		{\bibinfo  {journal} {Phys. Rev. Res.}\ }\textbf {\bibinfo {volume} {4}},\
		\bibinfo {pages} {023005} (\bibinfo {year} {2022})}\BibitemShut {NoStop}%
	\bibitem [{\citenamefont {Nocedal}\ and\ \citenamefont
		{Wright}(2006)}]{nocedal2006numerical}%
	\BibitemOpen
	\bibfield  {author} {\bibinfo {author} {\bibfnamefont {J.}~\bibnamefont
			{Nocedal}}\ and\ \bibinfo {author} {\bibfnamefont {S.~J.}\ \bibnamefont
			{Wright}},\ }\href@noop {} {\emph {\bibinfo {title} {Numerical
				Optimization}}},\ \bibinfo {edition} {2nd}\ ed.\ (\bibinfo  {publisher}
	{Springer},\ \bibinfo {address} {New York, NY},\ \bibinfo {year}
	{2006})\BibitemShut {NoStop}%
	\bibitem [{\citenamefont {Yuan}(2000)}]{yuan2000review}%
	\BibitemOpen
	\bibfield  {author} {\bibinfo {author} {\bibfnamefont {Y.-x.}\ \bibnamefont
			{Yuan}},\ }\bibfield  {title} {\bibinfo {title} {A review of trust region
			algorithms for optimization},\ }in\ \href
	{https://doi.org/10.1093/oso/9780198505143.003.0023} {\emph {\bibinfo
			{booktitle} {ICIAM99: Proceedings of the Fourth International Congress on
				Industrial \&amp; Applied Mathematics Edinburgh}}}\ (\bibinfo  {publisher}
	{Oxford University Press},\ \bibinfo {year} {2000})\ \Eprint
	{https://arxiv.org/abs/https://academic.oup.com/book/0/chapter/422191153/chapter-pdf/52438596/isbn-9780198505143-book-part-23.pdf}
	{https://academic.oup.com/book/0/chapter/422191153/chapter-pdf/52438596/isbn-9780198505143-book-part-23.pdf}
	\BibitemShut {NoStop}%
	\bibitem [{\citenamefont {Arnold}(1992)}]{ArnoldODE}%
	\BibitemOpen
	\bibfield  {author} {\bibinfo {author} {\bibfnamefont {V.~I.}\ \bibnamefont
			{Arnold}},\ }\href@noop {} {\emph {\bibinfo {title} {Ordinary Differential
				Equations}}}\ (\bibinfo  {publisher} {Springer-Verlag},\ \bibinfo {address}
	{Berlin},\ \bibinfo {year} {1992})\BibitemShut {NoStop}%
	\bibitem [{\citenamefont {G{\"u}nther}\ and\ \citenamefont
		{Rojo}(2021)}]{Gnther2021IntroductionTV}%
	\BibitemOpen
	\bibfield  {author} {\bibinfo {author} {\bibfnamefont {T.}~\bibnamefont
			{G{\"u}nther}}\ and\ \bibinfo {author} {\bibfnamefont {I.~B.}\ \bibnamefont
			{Rojo}},\ }\bibfield  {title} {\bibinfo {title} {Introduction to vector field
			topology},\ }\href {https://api.semanticscholar.org/CorpusID:227177128}
	{\bibfield  {journal} {\bibinfo  {journal} {Mathematics and Visualization}\ }
		(\bibinfo {year} {2021})}\BibitemShut {NoStop}%
	\bibitem [{\citenamefont {Aleksandrov}(2022)}]{universe8040223}%
	\BibitemOpen
	\bibfield  {author} {\bibinfo {author} {\bibfnamefont {A.~G.}\ \bibnamefont
			{Aleksandrov}},\ }\bibfield  {title} {\bibinfo {title} {The poincaré index
			and its applications},\ }\bibfield  {journal} {\bibinfo  {journal}
		{Universe}\ }\textbf {\bibinfo {volume} {8}},\ \href
	{https://doi.org/10.3390/universe8040223} {10.3390/universe8040223} (\bibinfo
	{year} {2022})\BibitemShut {NoStop}%
	\bibitem [{\citenamefont {Aad}\ \emph {et~al.}(2020)\citenamefont {Aad},
		\citenamefont {Abbott}, \citenamefont {Abbott}, \citenamefont {Abdinov},
		\citenamefont {Abed~Abud}, \citenamefont {Abeling}, \citenamefont
		{Abhayasinghe}, \citenamefont {Abidi}, \citenamefont {AbouZeid},
		\citenamefont {Abraham}, \citenamefont {Abramowicz}, \citenamefont {Abreu},
		\citenamefont {Abulaiti}, \citenamefont {Acharya}, \citenamefont {Achkar},
		\citenamefont {Adachi}, \citenamefont {Adam}, \citenamefont
		{Adam~Bourdarios}, \citenamefont {Adamczyk}, \citenamefont {Adamek},
		\citenamefont {Adelman}, \citenamefont {Adersberger}, \citenamefont
		{Adiguzel}, \citenamefont {Adorni}, \citenamefont {Adye}, \citenamefont
		{Affolder}, \citenamefont {Afik}, \citenamefont {Agapopoulou}, \citenamefont
		{Agaras}, \citenamefont {Aggarwal}, \citenamefont {Agheorghiesei},
		\citenamefont {Aguilar-Saavedra}, \citenamefont {Ahmadov}, \citenamefont
		{Ahmed}, \citenamefont {Ai}, \citenamefont {Aielli}, \citenamefont
		{Akatsuka},\ and\ \citenamefont {et~al.}}]{PhysRevLett.124.031802}%
	\BibitemOpen
	\bibfield  {author} {\bibinfo {author} {\bibfnamefont {G.}~\bibnamefont
			{Aad}}, \bibinfo {author} {\bibfnamefont {B.}~\bibnamefont {Abbott}},
		\bibinfo {author} {\bibfnamefont {D.~C.}\ \bibnamefont {Abbott}}, \bibinfo
		{author} {\bibfnamefont {O.}~\bibnamefont {Abdinov}}, \bibinfo {author}
		{\bibfnamefont {A.}~\bibnamefont {Abed~Abud}}, \bibinfo {author}
		{\bibfnamefont {K.}~\bibnamefont {Abeling}}, \bibinfo {author} {\bibfnamefont
			{D.~K.}\ \bibnamefont {Abhayasinghe}}, \bibinfo {author} {\bibfnamefont
			{S.~H.}\ \bibnamefont {Abidi}}, \bibinfo {author} {\bibfnamefont {O.~S.}\
			\bibnamefont {AbouZeid}}, \bibinfo {author} {\bibfnamefont {N.~L.}\
			\bibnamefont {Abraham}}, \bibinfo {author} {\bibfnamefont {H.}~\bibnamefont
			{Abramowicz}}, \bibinfo {author} {\bibfnamefont {H.}~\bibnamefont {Abreu}},
		\bibinfo {author} {\bibfnamefont {Y.}~\bibnamefont {Abulaiti}}, \bibinfo
		{author} {\bibfnamefont {B.~S.}\ \bibnamefont {Acharya}}, \bibinfo {author}
		{\bibfnamefont {B.}~\bibnamefont {Achkar}}, \bibinfo {author} {\bibfnamefont
			{S.}~\bibnamefont {Adachi}}, \bibinfo {author} {\bibfnamefont
			{L.}~\bibnamefont {Adam}}, \bibinfo {author} {\bibfnamefont {C.}~\bibnamefont
			{Adam~Bourdarios}}, \bibinfo {author} {\bibfnamefont {L.}~\bibnamefont
			{Adamczyk}}, \bibinfo {author} {\bibfnamefont {L.}~\bibnamefont {Adamek}},
		\bibinfo {author} {\bibfnamefont {J.}~\bibnamefont {Adelman}}, \bibinfo
		{author} {\bibfnamefont {M.}~\bibnamefont {Adersberger}}, \bibinfo {author}
		{\bibfnamefont {A.}~\bibnamefont {Adiguzel}}, \bibinfo {author}
		{\bibfnamefont {S.}~\bibnamefont {Adorni}}, \bibinfo {author} {\bibfnamefont
			{T.}~\bibnamefont {Adye}}, \bibinfo {author} {\bibfnamefont {A.~A.}\
			\bibnamefont {Affolder}}, \bibinfo {author} {\bibfnamefont {Y.}~\bibnamefont
			{Afik}}, \bibinfo {author} {\bibfnamefont {C.}~\bibnamefont {Agapopoulou}},
		\bibinfo {author} {\bibfnamefont {M.~N.}\ \bibnamefont {Agaras}}, \bibinfo
		{author} {\bibfnamefont {A.}~\bibnamefont {Aggarwal}}, \bibinfo {author}
		{\bibfnamefont {C.}~\bibnamefont {Agheorghiesei}}, \bibinfo {author}
		{\bibfnamefont {J.~A.}\ \bibnamefont {Aguilar-Saavedra}}, \bibinfo {author}
		{\bibfnamefont {F.}~\bibnamefont {Ahmadov}}, \bibinfo {author} {\bibfnamefont
			{W.~S.}\ \bibnamefont {Ahmed}}, \bibinfo {author} {\bibfnamefont
			{X.}~\bibnamefont {Ai}}, \bibinfo {author} {\bibfnamefont {G.}~\bibnamefont
			{Aielli}}, \bibinfo {author} {\bibfnamefont {S.}~\bibnamefont {Akatsuka}},\
		and\ \bibinfo {author} {\bibnamefont {et~al.}} (\bibinfo {collaboration}
		{ATLAS Collaboration}),\ }\bibfield  {title} {\bibinfo {title} {Search for
			magnetic monopoles and stable high-electric-charge objects in 13 tev
			proton-proton collisions with the atlas detector},\ }\href
	{https://doi.org/10.1103/PhysRevLett.124.031802} {\bibfield  {journal}
		{\bibinfo  {journal} {Phys. Rev. Lett.}\ }\textbf {\bibinfo {volume} {124}},\
		\bibinfo {pages} {031802} (\bibinfo {year} {2020})}\BibitemShut {NoStop}%
	\bibitem [{\citenamefont {Abo-Shaeer}\ \emph {et~al.}(2001)\citenamefont
		{Abo-Shaeer}, \citenamefont {Raman}, \citenamefont {Vogels},\ and\
		\citenamefont {Ketterle}}]{doi:10.1126/science.1060182}%
	\BibitemOpen
	\bibfield  {author} {\bibinfo {author} {\bibfnamefont {J.~R.}\ \bibnamefont
			{Abo-Shaeer}}, \bibinfo {author} {\bibfnamefont {C.}~\bibnamefont {Raman}},
		\bibinfo {author} {\bibfnamefont {J.~M.}\ \bibnamefont {Vogels}},\ and\
		\bibinfo {author} {\bibfnamefont {W.}~\bibnamefont {Ketterle}},\ }\bibfield
	{title} {\bibinfo {title} {Observation of vortex lattices in bose-einstein
			condensates},\ }\href {https://doi.org/10.1126/science.1060182} {\bibfield
		{journal} {\bibinfo  {journal} {Science}\ }\textbf {\bibinfo {volume}
			{292}},\ \bibinfo {pages} {476} (\bibinfo {year} {2001})},\ \Eprint
	{https://arxiv.org/abs/https://www.science.org/doi/pdf/10.1126/science.1060182}
	{https://www.science.org/doi/pdf/10.1126/science.1060182} \BibitemShut
	{NoStop}%
	\bibitem [{\citenamefont {Zwierlein}\ \emph {et~al.}(2005)\citenamefont
		{Zwierlein}, \citenamefont {Abo-Shaeer}, \citenamefont {Abo-Shaeer},
		\citenamefont {Schirotzek}, \citenamefont {Schunck},\ and\ \citenamefont
		{Ketterle}}]{Zwierlein2005VorticesAS}%
	\BibitemOpen
	\bibfield  {author} {\bibinfo {author} {\bibfnamefont {M.~W.}\ \bibnamefont
			{Zwierlein}}, \bibinfo {author} {\bibfnamefont {J.~R.}\ \bibnamefont
			{Abo-Shaeer}}, \bibinfo {author} {\bibfnamefont {J.~R.}\ \bibnamefont
			{Abo-Shaeer}}, \bibinfo {author} {\bibfnamefont {A.}~\bibnamefont
			{Schirotzek}}, \bibinfo {author} {\bibfnamefont {C.~H.}\ \bibnamefont
			{Schunck}},\ and\ \bibinfo {author} {\bibfnamefont {W.}~\bibnamefont
			{Ketterle}},\ }\bibfield  {title} {\bibinfo {title} {Vortices and
			superfluidity in a strongly interacting fermi gas},\ }\href
	{https://doi.org/10.1038/nature03858} {\bibfield  {journal} {\bibinfo
			{journal} {Nature}\ }\textbf {\bibinfo {volume} {435}},\ \bibinfo {pages}
		{1047} (\bibinfo {year} {2005})}\BibitemShut {NoStop}%
	\bibitem [{\citenamefont {Wolf}\ \emph {et~al.}(2022)\citenamefont {Wolf},
		\citenamefont {Boegel}, \citenamefont {Meister}, \citenamefont
		{Bala\ifmmode~\check{z}\else \v{z}\fi{}}, \citenamefont {Gaaloul},\ and\
		\citenamefont {Efremov}}]{PhysRevA.106.013309}%
	\BibitemOpen
	\bibfield  {author} {\bibinfo {author} {\bibfnamefont {A.}~\bibnamefont
			{Wolf}}, \bibinfo {author} {\bibfnamefont {P.}~\bibnamefont {Boegel}},
		\bibinfo {author} {\bibfnamefont {M.}~\bibnamefont {Meister}}, \bibinfo
		{author} {\bibfnamefont {A.}~\bibnamefont {Bala\ifmmode~\check{z}\else
				\v{z}\fi{}}}, \bibinfo {author} {\bibfnamefont {N.}~\bibnamefont {Gaaloul}},\
		and\ \bibinfo {author} {\bibfnamefont {M.~A.}\ \bibnamefont {Efremov}},\
	}\bibfield  {title} {\bibinfo {title} {Shell-shaped bose-einstein condensates
			based on dual-species mixtures},\ }\href
	{https://doi.org/10.1103/PhysRevA.106.013309} {\bibfield  {journal} {\bibinfo
			{journal} {Phys. Rev. A}\ }\textbf {\bibinfo {volume} {106}},\ \bibinfo
		{pages} {013309} (\bibinfo {year} {2022})}\BibitemShut {NoStop}%
	\bibitem [{\citenamefont {Pethick}\ and\ \citenamefont
		{Smith}(2008)}]{Pethick-BEC}%
	\BibitemOpen
	\bibfield  {author} {\bibinfo {author} {\bibfnamefont {C.~J.}\ \bibnamefont
			{Pethick}}\ and\ \bibinfo {author} {\bibfnamefont {H.}~\bibnamefont
			{Smith}},\ }\href {https://doi.org/10.1017/CBO9780511802850} {\emph {\bibinfo
			{title} {Bose–Einstein Condensation in Dilute Gases}}},\ \bibinfo {edition}
	{2nd}\ ed.\ (\bibinfo  {publisher} {Cambridge University Press},\ \bibinfo
	{address} {Cambridge, UK},\ \bibinfo {year} {2008})\BibitemShut {NoStop}%
	\bibitem [{\citenamefont {Chien}\ \emph {et~al.}(2006)\citenamefont {Chien},
		\citenamefont {He}, \citenamefont {Chen},\ and\ \citenamefont
		{Levin}}]{PhysRevA.73.041603}%
	\BibitemOpen
	\bibfield  {author} {\bibinfo {author} {\bibfnamefont {C.-C.}\ \bibnamefont
			{Chien}}, \bibinfo {author} {\bibfnamefont {Y.}~\bibnamefont {He}}, \bibinfo
		{author} {\bibfnamefont {Q.}~\bibnamefont {Chen}},\ and\ \bibinfo {author}
		{\bibfnamefont {K.}~\bibnamefont {Levin}},\ }\bibfield  {title} {\bibinfo
		{title} {Ground-state description of a single vortex in an atomic fermi gas:
			From bcs to bose--einstein condensation},\ }\href
	{https://doi.org/10.1103/PhysRevA.73.041603} {\bibfield  {journal} {\bibinfo
			{journal} {Phys. Rev. A}\ }\textbf {\bibinfo {volume} {73}},\ \bibinfo
		{pages} {041603} (\bibinfo {year} {2006})}\BibitemShut {NoStop}%
	\bibitem [{\citenamefont {P\'erez-Garrido}\ \emph {et~al.}(1997)\citenamefont
		{P\'erez-Garrido}, \citenamefont {Dodgson},\ and\ \citenamefont
		{Moore}}]{PhysRevB.56.3640}%
	\BibitemOpen
	\bibfield  {author} {\bibinfo {author} {\bibfnamefont {A.}~\bibnamefont
			{P\'erez-Garrido}}, \bibinfo {author} {\bibfnamefont {M.~J.~W.}\ \bibnamefont
			{Dodgson}},\ and\ \bibinfo {author} {\bibfnamefont {M.~A.}\ \bibnamefont
			{Moore}},\ }\bibfield  {title} {\bibinfo {title} {Influence of dislocations
			in thomson's problem},\ }\href {https://doi.org/10.1103/PhysRevB.56.3640}
	{\bibfield  {journal} {\bibinfo  {journal} {Phys. Rev. B}\ }\textbf {\bibinfo
			{volume} {56}},\ \bibinfo {pages} {3640} (\bibinfo {year}
		{1997})}\BibitemShut {NoStop}%
\end{thebibliography}
%

\end{document}